\documentclass[12pt,A4paper]{article}

\usepackage{amsmath}
\usepackage{epsfig,amssymb}
\usepackage{graphicx}
\usepackage[latin1]{inputenc}       
\newcommand{\om}{\omega}

\usepackage{color}

\begin{document}

\title{\bf Thermal noise properties of two aging materials  }
\author{  L. Bellon, L. Buisson, M. Ciccotti, \underline{S. Ciliberto $^\diamondsuit$},  F.
Douarche \\
Ecole Normale Supérieure de Lyon, Laboratoire de Physique,\\
C.N.R.S. UMR5672,  \\ 46, All\'ee d'Italie, 69364 Lyon Cedex
07,  France\\
\\
{\noindent{\it $^\diamondsuit$ {correspondence to:
Sergio.Ciliberto@ens-lyon.fr}}}
 }
 \date{}
 \maketitle

\begin{abstract}
In this lecture we review several aspects of the thermal noise
properties in  two aging materials:   a polymer and a colloidal
glass.
 The measurements have been performed     after a quench for the polymer and during the transition
 from a fluid-like to a solid-like state for the gel. Two kind of
 noise has been measured: the electrical noise and the mechanical
 noise.
  For both materials we have observed that the electric noise is characterized by a strong
   intermittency, which induces a large violation of the Fluctuation Dissipation Theorem (FDT)
    during the aging time, and may persist for several hours at low frequency.
    The statistics of these intermittent signals and their dependance
    on the quench speed for the polymer or on sample concentration for the gel are studied.
    The results are in a qualitative agreement with recent models of aging, that predict an intermittent dynamics.
For the mechanical noise the results are unclear. In the polymer
the mechanical thermal noise is still intermittent whereas for the
gel the violation of FDT, if it exists, is extremely small.
\end{abstract}

\section{Introduction}
When a glassy system is quenched from above to below the glass
transition temperature $T_g$, any response function of the
material depends on the time $t_w$ elapsed from the
quench\cite{Struick}. For example, the dielectric and elastic
constants of polymers continue to evolve several years after the
quench \cite{Struick}.   Similarly,   the magnetic susceptibility
of spin-glasses depends on the time spent at low temperature
\cite{book}. Another example of aging is given by
colloidal-glasses, whose properties evolve during the sol-gel
transition which may last a few  days \cite{Kroon}.  For obvious
reasons related to applications, aging has been mainly
characterized by the study of the slow time evolution of response
functions, such as the dielectric and elastic properties of these
materials. It has been observed that these systems may present
very complex effects, such as memory and
rejuvenation\cite{Struick,Kovacs,Jonason,bellonM}, in other words
their physical properties depend on the   whole   thermal history
of the sample. Many models and theories have been constructed in
order to explain the observed phenomenology, which is not yet
completely understood. These models either predict or assume very
different dynamical behaviours of the systems during aging. This
dynamical behaviour can be directly related to the thermal noise
features of these aging systems and the study of response
functions alone is unable to give definitive   answers on the
approaches   that are the most adapted to explain the aging of a
specific material. Thus it is important to associate the measure
of thermal noise to that of response functions. The measurement of
fluctuations is also related to another important aspect of aging
dynamics. Indeed glasses are out of equilibrium systems  and usual
thermodynamics does not necessarily apply. However, as the time
evolution is slow, some concepts of the classical approach may be
useful for understanding the glass aging properties. A widely
studied question is the definition of an effective temperature in
these systems which are weakly, but durably, out of equilibrium.
Recent theories\cite{Kurchan} based on the description of spin
glasses by a mean field approach proposed to extend the concept of
temperature using a Fluctuation Dissipation Relation (FDR) which
generalizes the Fluctuation Dissipation Theorem (FDT) for a weakly
out of equilibrium system (for a review see
Ref.\cite{Mezard,Cugliandolo,Peliti}). However the validity of
this temperature is still an open and widely studied question.

 For all of these reasons,
in recent years, the study of the thermal noise of aging materials
has received a growing interest. However in spite of the large
amount of theoretical studies there are only a few experiments
dedicated to this problem \cite{Grigera}-\cite{Cipelletti}. The
available experimental results are in some way in contradiction
and they are unable to give definitive answers. For example the
thermal noise may present a strong intermittency which slowly
disappear during aging. Although several theoretical models
predict this intermittency \cite{Sollich,Miguel1,Miguel2,Sibani}
the experimental conditions which produce such a kind of behaviour
are unclear.  Therefore new experiments are necessary to increase
our knowledge on the thermal noise properties of the aging
materials.

In this lecture we  will review several experimental results on
the electrical and mechanical thermal fluctuations  of a polymer
  and a colloidal glass.   We will mainly  focus on     the
measurements of the dielectric susceptibility and of the
polarization noise   in the polymer material , in the range
$20mHz-100Hz$, because the results of these measurements
demonstrate the appearance of a strong intermittency of the noise
when this material is quickly quenched from the molten state to
below its glass-transition temperature.  This intermittency
produces a strong violation of the FDT at very low frequency. The
violation is a decreasing function of the time  $t_w$ elapsed from
the quench, and of the frequency of measurement $f = \omega / 2
\pi$. Nevertheless, this violation   is observed at $\omega t_w
\gg 1$ and      may last for more than $3h$ for $f>1Hz$. We have
also observed that the intermittency is a function of the cooling
rate of the sample and it almost disappears after a slow quench.
In this case the violation of FDT remains but it is very small.
Preliminary mechanical measurements done on a polycarbonate beam
confirm the presence of an intermittent behaviour after a fast
quench.   We also review some equivalent measurements in a
different material: a colloidal glass of Laponite. As for the
polymer, a strong intermittency, sensible to initial conditions,
is observed with electrical measurements. It is interesting to
note however that no such effect can be detected on the mechanical
behaviour.

The paper is organized in three main sections, the
first one on the electrical measurements in polycarbonate, a
second one on the mechanical measurements and a third one on the
  fluctuations in Laponite preparations. In the first section we
 describe the experimental set up and the measurement procedure,
  the results of the noise and response measurements and the
statistical analysis of the noise. We then discuss the dependence
on the quench speed of the FDT violation and  the temporal
behaviour of the effective temperature after a slow quench. In
section 3  we describe the experimental set-up for mechanical
measurements and the preliminary results on the mechanical noise.
In section 4 the electrical noise in   the   colloidal gel is
analyzed. We briefly   discuss   the case of the mechanical
properties for the gel. Finally in section 5 we first compare the
experimental results on polycarbonate with those of colloidal
glasses and other materials. We then discuss the relevance of
these results in the context of the recent theoretical models
before concluding.

\section{ Dielectric noise in a polymer glass}

We present in this section measurements of the dielectric
susceptibility and of the polarization noise, in the range
$20\,mHz\,-\,100\,Hz$, of a polymer glass: polycarbonate. These
results demonstrate the appearance of a strong intermittency of
the noise when this material is quickly quenched from the molten
state to below its glass-transition temperature. This
intermittency produces a strong violation of the FDT at very low
frequency. The violation is a decreasing function of time and
frequency and it is still observed for $\omega t_w \gg 1$: it may
last for more than $3h$ for $f>1\,Hz$. We have also observed that
the intermittency is a function of the cooling rate of the sample
and almost disappears after a slow quench. In this case the
violation of FDT remains, but it is very small.

\subsection{ Experimental setup}

The polymer used in this investigation is Makrofol DE 1-1 C, a
bisphenol A polycarbonate, with $T_g \simeq 419K$, produced by
Bayer in form of foils. We have chosen this material because it
has a wide temperature range of strong aging\cite{Struick}. This
polymer is totally amorphous: there is no evidence of
crystallinity\cite{Wilkes1}. Nevertheless, the internal structure
of polycarbonate changes and relaxes as a result of a change in
the chain conformation by molecular
motions\cite{Struick},\cite{Duval},\cite{Quinson}. Many studies of
the dielectric susceptibility of this material exist, but none had
an interest on the problem of noise measurements.

\begin{figure}[!ht]
\begin{center}
\includegraphics[width=8cm]{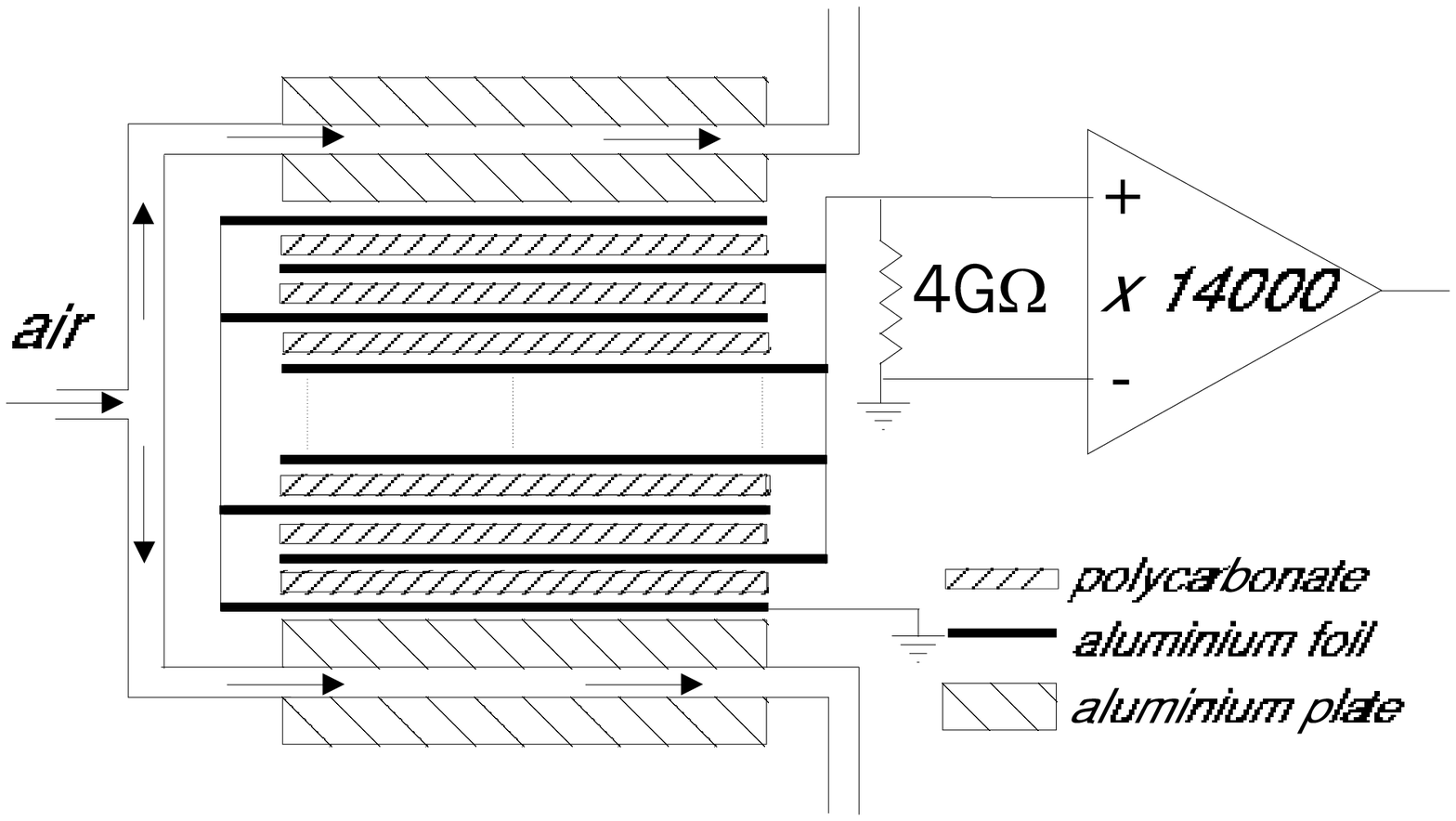}
\includegraphics[width=5cm]{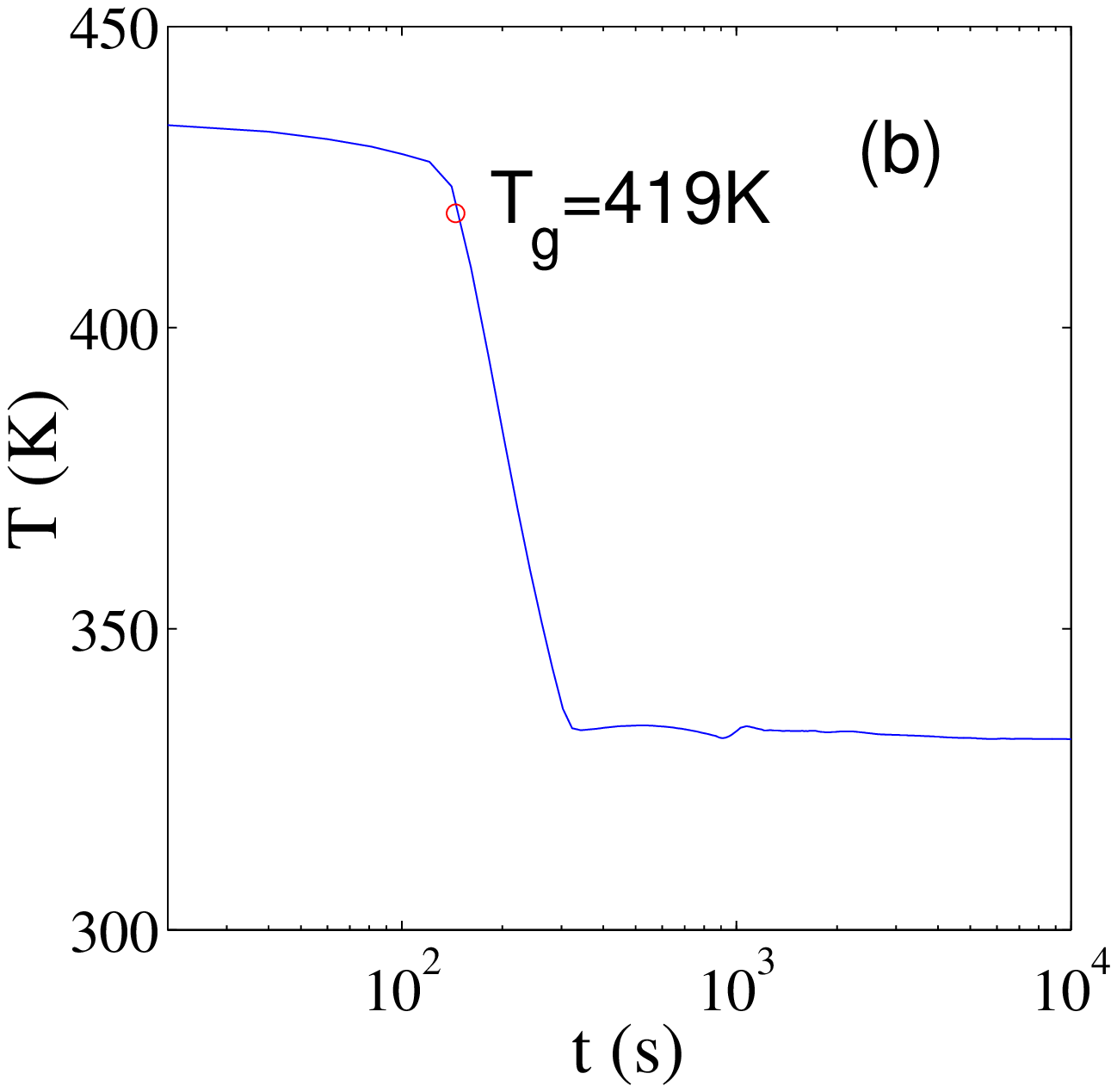}
\end{center}
\caption{{\bf  Polycarbonate experimental set-up}(a) Design of
polycarbonate capacitance cell. (b) Typical temperature quench:
from $T_i=453\,K$ to $T_f=333\,K$, the origin of $t_w$ is set at
$T=T_g$. }
 \label{Experimental set-up}
\end{figure}

In our experiment polycarbonate is used as the dielectric of a
capacitor. The capacitor is composed by $14$ cylindrical
capacitors in parallel in order to reduce the resistance of the
sample and to increase its capacity \cite{BuissonD}. Each
capacitor is made of two aluminum electrodes, $12\mu m$ thick, and
by a disk of polycarbonate of diameter $12cm$ and thickness
$125\mu m$. The experimental set-up is shown in
Fig.~~\ref{Experimental set-up}(a). The $14$ capacitors are
sandwiched together and put inside two thick aluminum plates which
contain an air circulation used to regulate the sample
temperature. This mechanical design of the capacitor is very
stable and gives very reproducible results even after many
temperature quenches. The capacitor is inside 4 Faraday screens to
insulate it from external noise. The temperature of the sample is
controlled within a few percent. Fast   quenches   of about $1K/s$
are obtained by injecting Nitrogen vapor in the air circulation of
the aluminum plates. The electrical impedance of the capacitor is
$Z(\omega,t_w) = R / (1+i \omega \ R \ C)$, where $C$ is the
capacitance and $R$ is a parallel resistance which accounts for
the complex dielectric susceptibility. This is measured by a
lock-in amplifier associated with an impedance adapter
 \cite{BuissonD}.   The noise spectrum $S_Z(\omega,t_w)$ of
the impedance $Z(\omega,t_w)$ is:
\begin{equation}
S_Z(f,t_w)= 4 \ k_B \ T_{eff}(f, t_w) \ Re [Z(\omega,t_w)]= {4 \
k_B \ T_{eff}(f, t_w) \ R \over 1+ (\omega \ R \ C)^2 } \label{SZ}
\end{equation}
where $k_B$ is the Boltzmann constant and $T_{eff}$ is the
effective temperature of the sample. This effective temperature
takes into account the  fact that FDT     (Nyquist relation for
electric noise) can be violated because the polymer is out of
equilibrium during aging, and in general $T_{eff}>T$, with $T$ the
temperature of the thermal bath. Of course when FDT is satisfied
then $T_{eff}=T$. In order to measure $S_Z(f,t_w)$, we have made a
differential amplifier based on selected low noise JFET(2N6453
InterFET Corporation), whose input has been polarized by a
resistance $R_i= 4G\Omega$. Above $2Hz$, the input voltage noise
of this amplifier is $5nV/\sqrt{Hz}$ and the input current noise
is about $1fA/\sqrt{Hz}$. The output signal of the amplifier is
directly acquired by a NI4462 card. It is easy to show that the
measured spectrum at the amplifier input is:
\begin{eqnarray}
S_V(f,t_w)& = &{4 \ k_B \ R \ R_i \ \ (\ T_{eff}(f, t_w) \ R_i + \
T_R \ R + S_\xi(f) \ R \ R_i ) \over (R+R_i)^2+(\omega \ R \ R_i \
C)^2} + S_{\eta}(f)
 \label{Vnoise}
\end{eqnarray}
where $T_R$ is the temperature of $R_i$ and $S_\eta$ and $S_\xi$
are respectively the voltage and the current noise spectrum of the
amplifier. In order to reach the desired statistical accuracy of
$S_V(f,t_w)$, we averaged the results of many experiments. In each
of these experiments the sample is first heated to $T_i=1.08T_g$.
It is maintained at this temperature for several hours in order to
reinitialize its thermal history. Then it is quenched from $T_i$
to the working final temperature $T_f$ where the aging properties
are studied. The maximum quenching rate from $T_i$ to $T_f$ is
$1K/s$. A typical thermal history of a fast quench is shown in
Fig.~\ref{Experimental set-up}(b). The reproducibility of the
capacitor impedance, during this thermal cycle is always better
than $1\%$. The origin of aging time $t_w$ is the instant when the
capacitor temperature is at $T_g \simeq 419 K$, which of course
may depend on the cooling rate. However adjustment of $T_g$ of a
few degrees will shift the time axis by at most $30s$, without
affecting our results.

\subsection{Response and noise measurements}

\begin{figure}[!ht]
\centerline{\hspace{1cm} \bf (a) \hspace{6cm} (b) }
\begin{center}
\includegraphics[width=7cm]{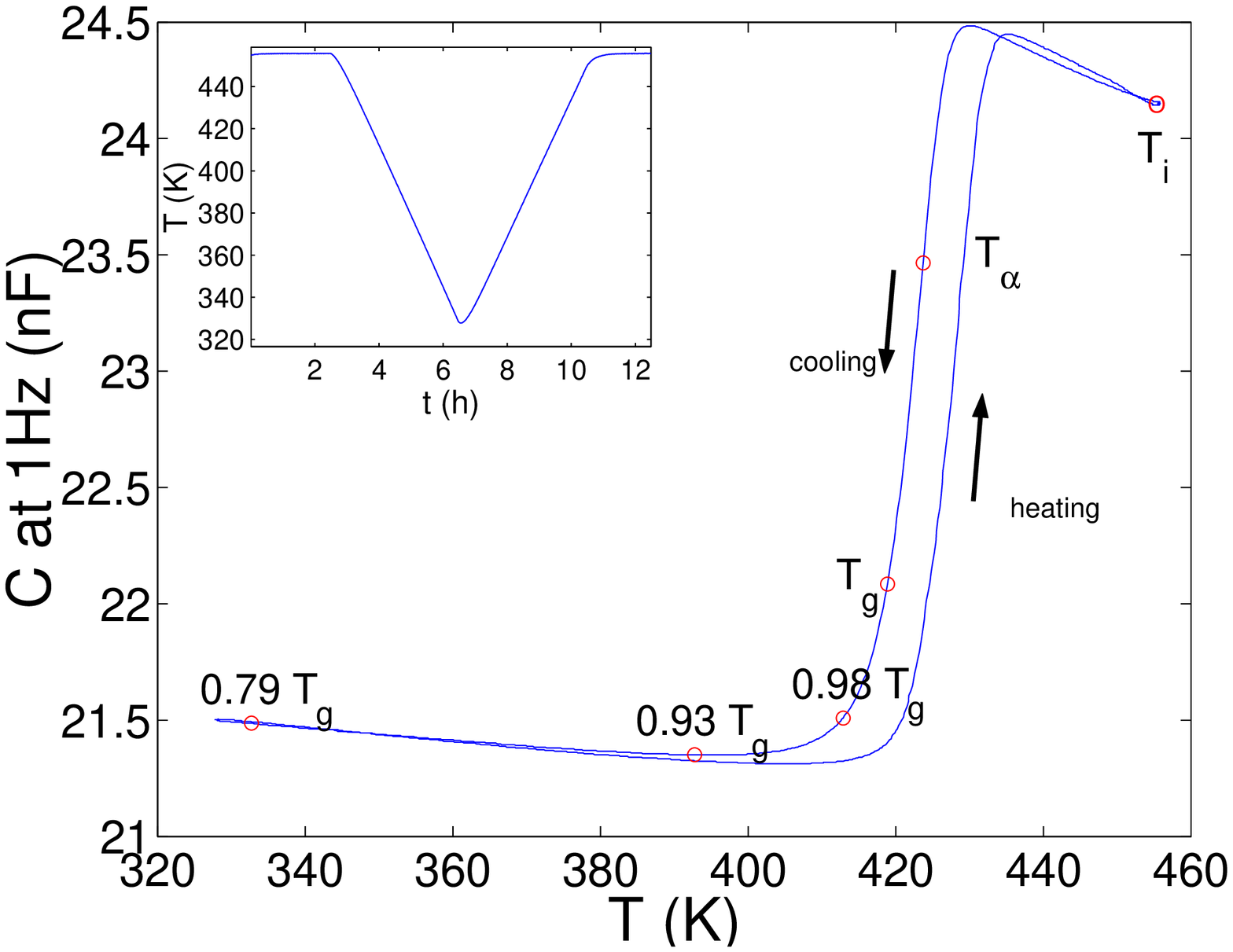}
\includegraphics[width=7cm]{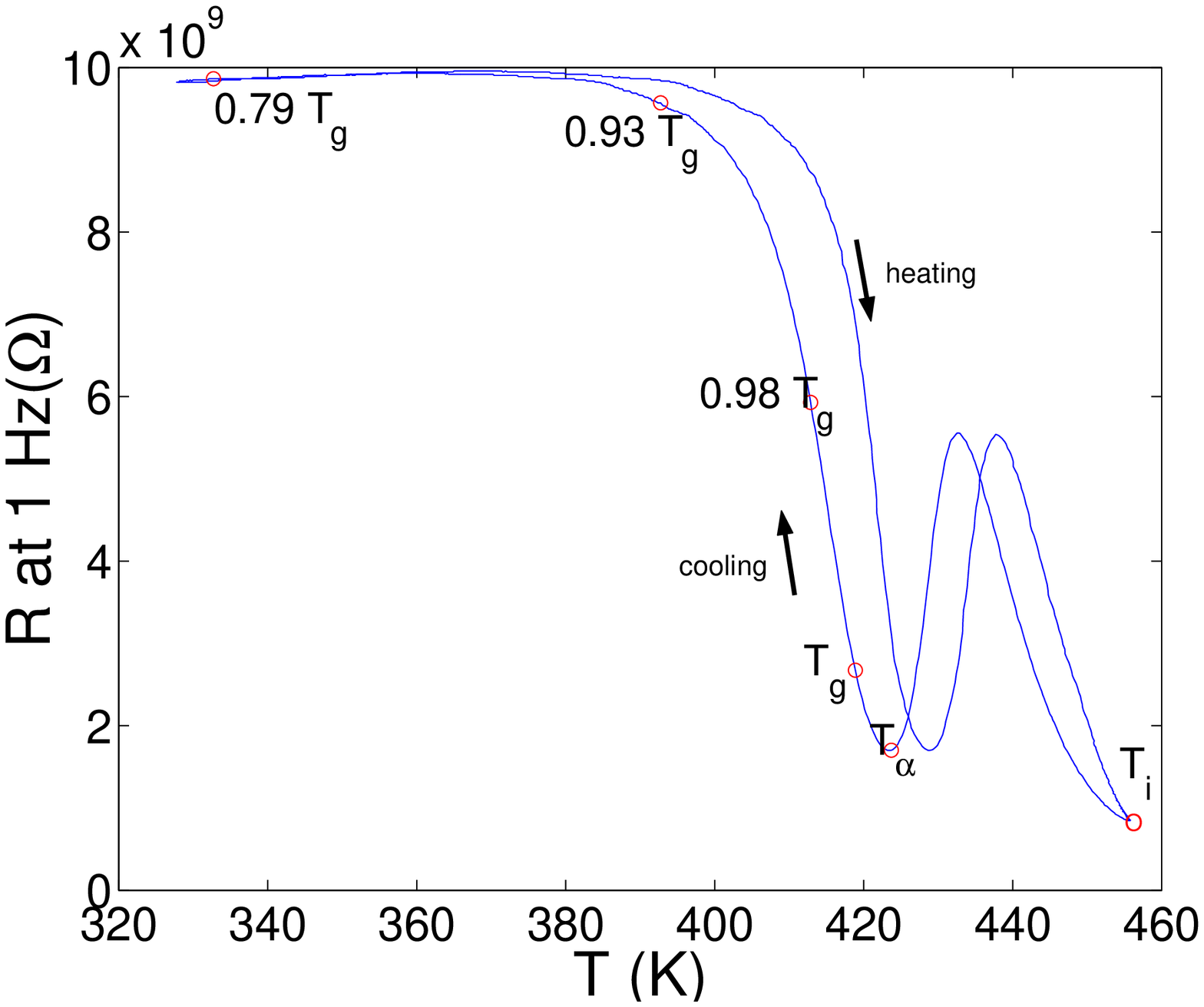}
\end{center}
\caption{{\bf Polycarbonate response function at 1Hz}(a)
Dependence of $C$, measured at $1Hz$, on temperature, when $T$ is
changed as function of time as indicated in the inset. (b)
Dependence of $R$, measured at $1Hz$, on $T$. $T_\alpha$ is the
temperature of the $\alpha$ relaxation at$1Hz$, $T_g$ is the glass
transition temperature. The other circles on the curve indicate
the $T_f$ where aging has been studied.} \label{hist}
\end{figure}

Before discussing the time evolution of the dielectric properties
and of the thermal noise at $T_f$  we show in Fig.~\ref{hist} the
dependence of $R$ and $C$ measured at $1Hz$ as a function of
temperature, which is ramped as a function of time as indicated in
the inset of Fig.~\ref{hist}(a).  We notice a strong hysteresis
between cooling and heating. In the figure $T_\alpha$ is the
temperature of the $\alpha$ relaxation at $1Hz$. The other circles
on the curve indicate the $T_f$ where the aging has been studied.
We have performed measurements at $T_f=0.79T_g,0.93T_g,0.98T_g$
using fast and slow quenches. The cooling rate is $1K/s$ and
$0.06K/s$ for the fast and slow quenches respectively.  As at
$T_f=0.98T_g$ the dielectric constant strongly depends on
temperature (see Fig.\ref{hist}), the temperature stability has to
be much better at $T_f=0.98T_g$ than at the two other smaller
$T_f$. Because of this good temperature stability needed at
$T_f=0.98T_g$ it is impossible to reach this temperature too fast.
Therefore at 0.98 $T_g$ we have performed only measurements after
a slow quench.

\begin{figure}[!ht]
\begin{center}
\includegraphics[width=10cm]{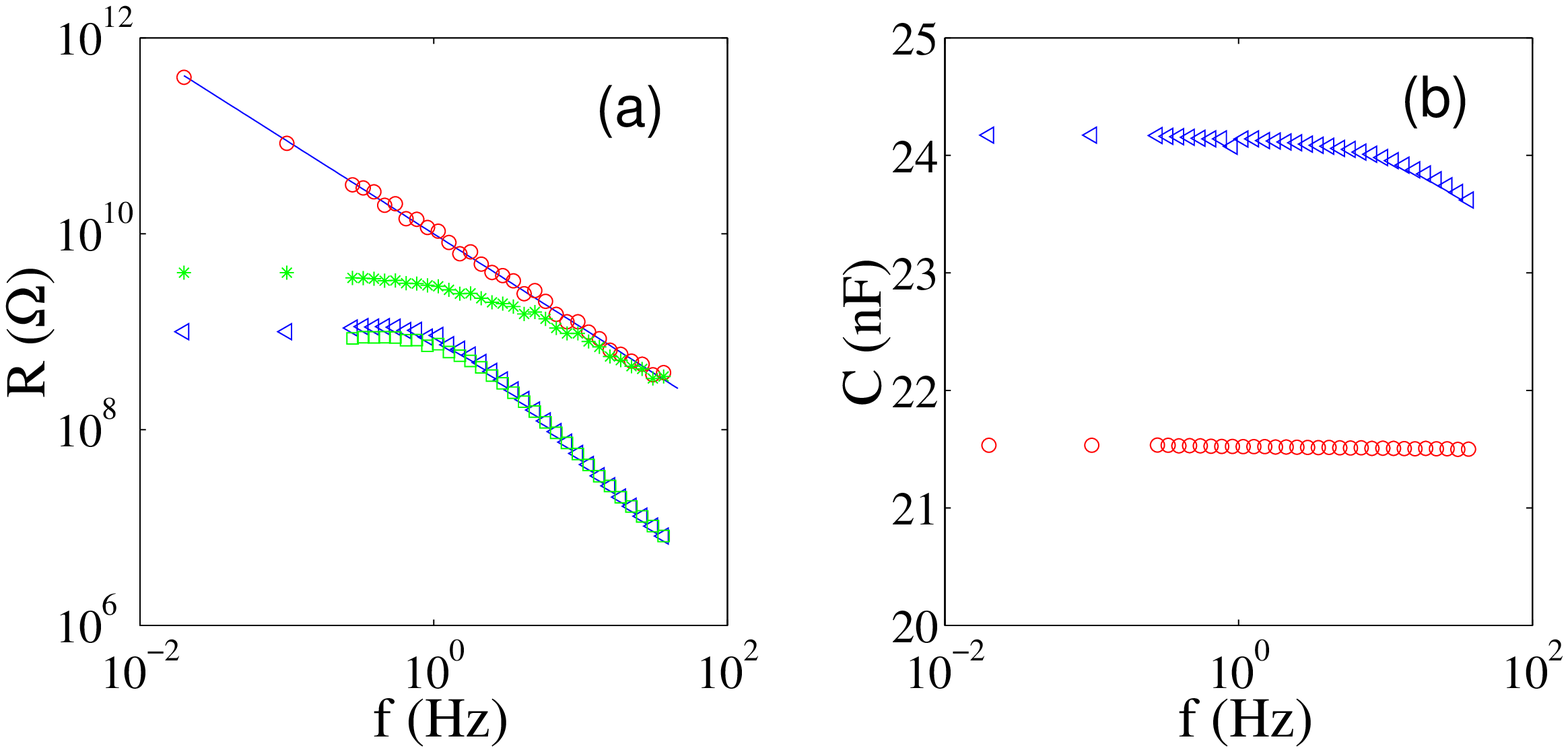}
\includegraphics[width=5cm]{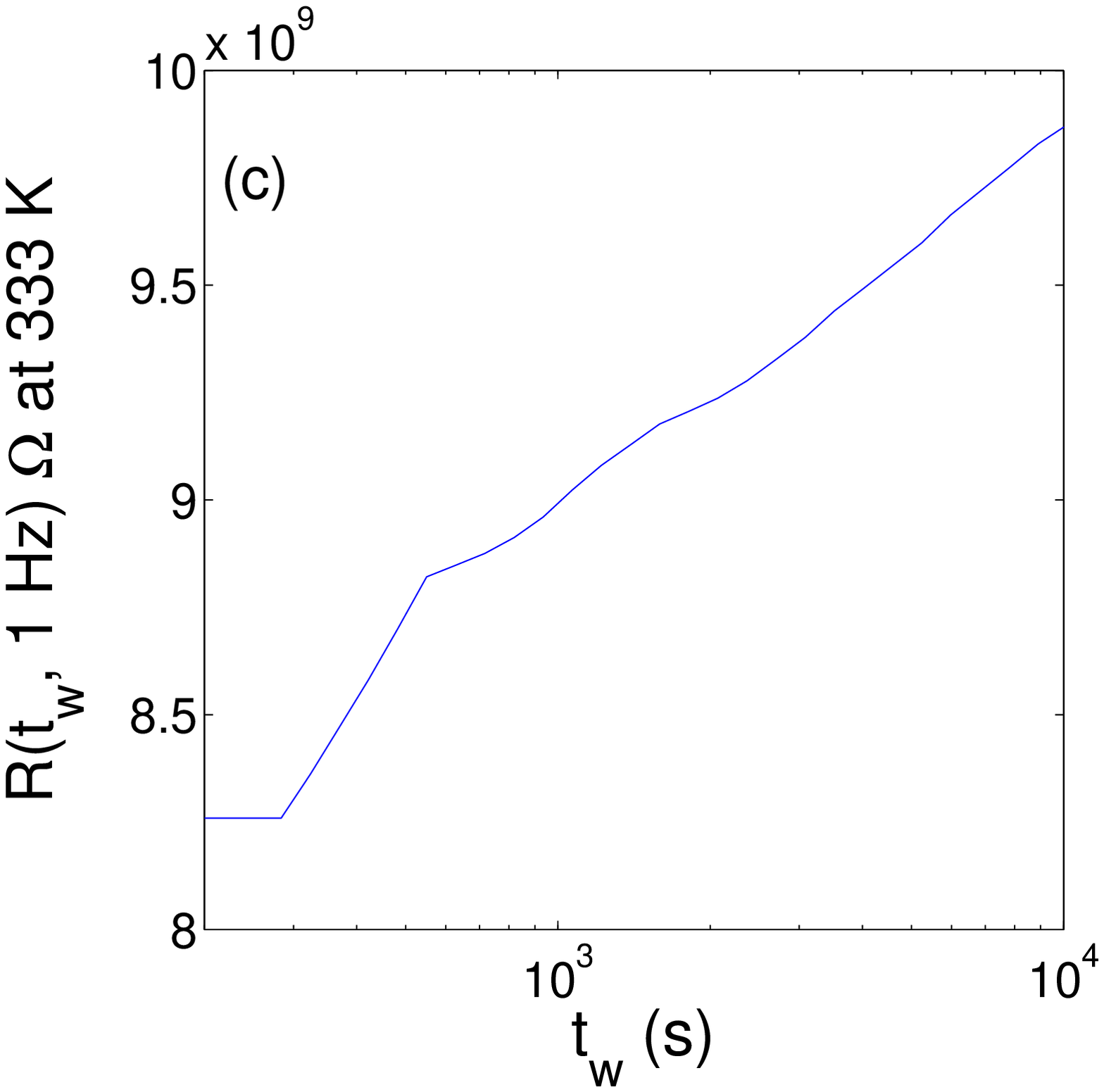}
\end{center}
\caption{{\bf Polycarbonate response function}(a) Polycarbonate
resistance $R$ as a function of frequency measured at
$T_i=1.08T_g$ ($\vartriangleleft$) and at $T_f=0.79T_g$
($\circ$)(after a fast quench). The effect of the $4G\Omega$ input
resistance in parallel with the polycarbonate impedance is also
shown at $T=433K$ ($\square$) and at $T=333K$ ($\ast$). (b)
Polycarbonate capacitance versus frequency measured at $T_i =433K$
($\vartriangleleft$) and at $T_f=333K$ ($\circ$). (c) Typical
aging of $R$ measured at $1Hz$ as a function of $t_w$}
\label{reponse}
\end{figure}

We first describe the results after a  fast quench at the smallest
temperature, that is $T_f=0.79T_g$. In Fig.~\ref{reponse}(a) and
(b), we plot the measured values of $R$ and $C$ as a function of
$f$ at $T_i=1.08T_g$ and at $T_f$ for $t_w \geqslant 200s$. The
dependence of $R$, at $1Hz$, as a function of time is shown in
Fig.~\ref{reponse}(c). We see that the time evolution of $R$ is
logarithmic in time for $t>300s$ and that the aging is not very
large at $T_f=0.79T_g$, it is only $10\%$ in 3 hours. At higher
temperature close to $T_g$ aging is much larger.

Looking at Fig.~\ref{reponse}(a) and (b), we see that lowering
temperature $R$ increases and $C$ decreases. As at $0.79T_g$ aging
is small and extremely slow for $t_w>200s$ the impedance can be
considered constant without affecting our results. From the data
plotted in Fig.~\ref{reponse} (a) and (b) one finds that
$R=10^{10}(1 \pm 0.05) \ f^{-1.05\pm 0.01} \ \Omega$ and $C=(21.5
\pm 0.05) nF$. In Fig.~\ref{reponse}(a) we also plot the total
resistance at the amplifier input which is the parallel of the
capacitor impedance with $R_i$. We see that at $T_f$ the input
impedance of the amplifier is negligible for $f>10Hz$, whereas it
has to be taken into account at slower frequencies.

\begin{figure}
\begin{center}
\includegraphics[width=7cm]{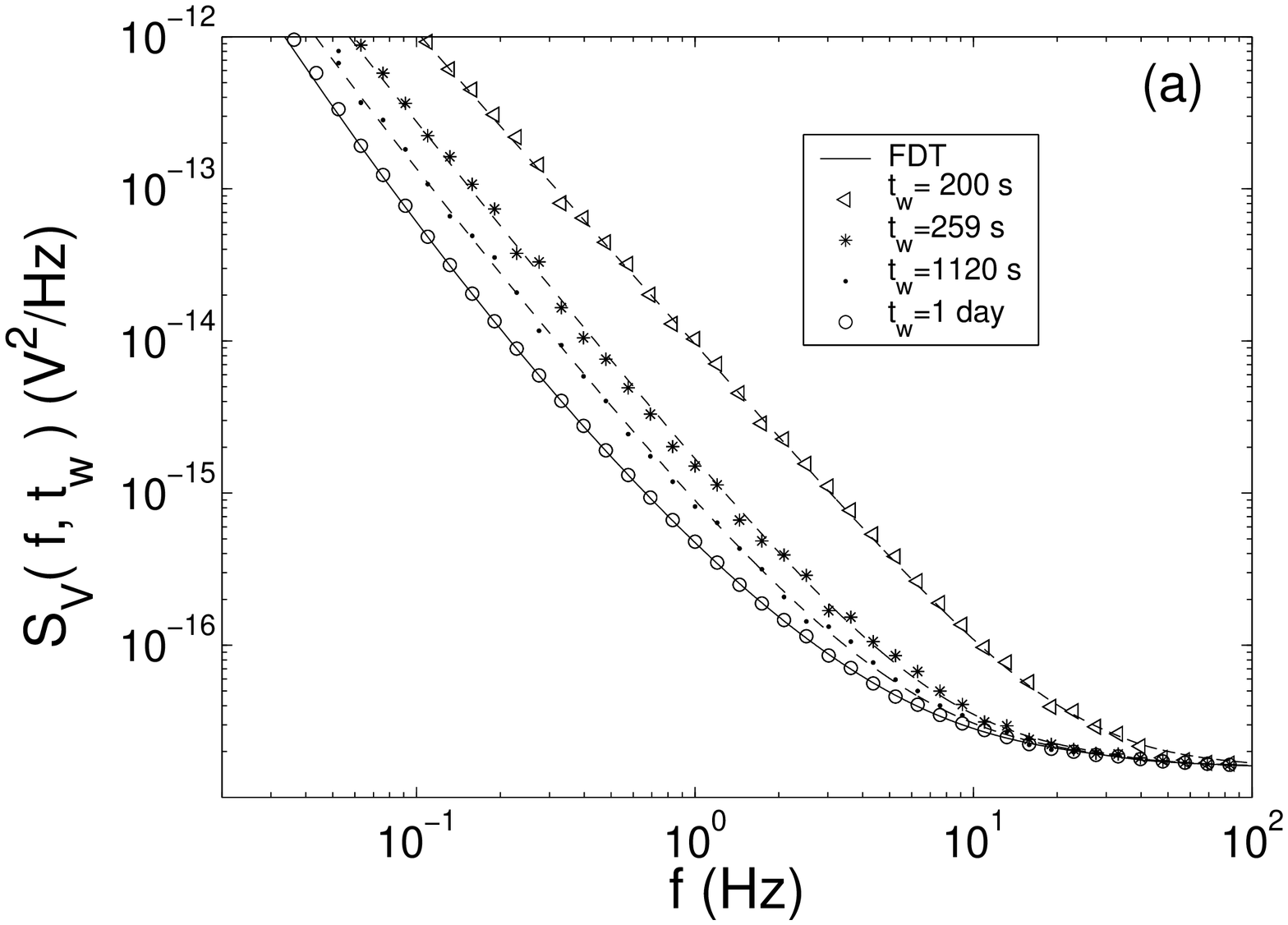}
\includegraphics[width=7cm]{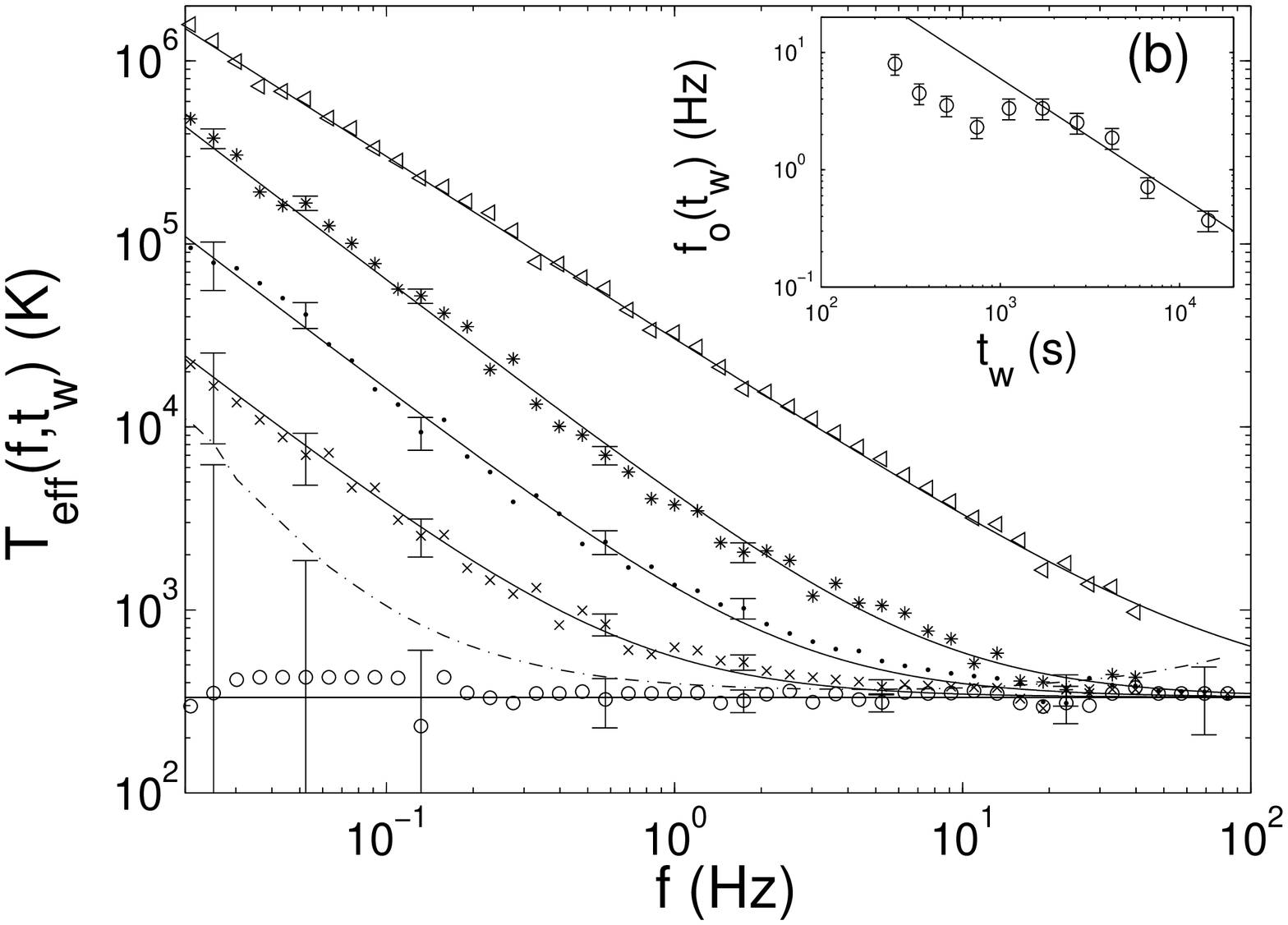}
\end{center}
\caption{{\bf Voltage noise and effective temperature in
polycarbonate after a fast quench}(a) Noise power spectral density
$S_V(f,t_w)$ measured at $T_f= 333K$ and different $t_w$. The
spectra are the average over seven quenches. The continuous line
is the FDT prediction. Dashed lines are the fit obtained using
eq.\ref{Vnoise} and eq.\ref{fitTeff} (see  text for details). (b)
Effective temperature vs frequency at $T_f=333K$ for different
aging times: $ (\vartriangleleft)\ tw= 200 \ s$, $ (\ast)\ tw= 260
s$, $ \bullet \ tw= 2580 s$, $ (\times) t_w=6542s$, $ (\circ)
t_w=1\ day $. The continuous lines are the fits obtained using
eq.\ref{fitTeff}. The horizontal straight line is the FDT
prediction. The dot dashed line corresponds to the limit where the
FDT violation can be detected. In the inset the frequency
$f_o(t_w)$, defined in eq.\ref{fitTeff},is plotted as a function
of $t_w$. The continuous line is not a fit, but it corresponds to
 $f_o(t_w) \propto 1/t_w$. }
\label{noise}
\end{figure}

Fig.~\ref{noise}(a) represents the evolution of $S_V(f,t_w)$ after
the fast  quench. Each spectrum is obtained as an average in a
time window starting at $t_w$. The time window increases with
$t_w$ so to reduce error for large $t_w$. The results of 7
quenches have been averaged. At the longest time ($t_w=1 \ day$)
the equilibrium FDT prediction (continuous line) is quite well
satisfied. We clearly see that FDT is strongly violated for all
frequencies at short times. Then high frequencies relax on the
FDT, but there is a persistence of the violation for lower
frequencies. The amount of the violation can be estimated by the
best fit of $T_{eff}(f,t_w)$ in eq.\ref{Vnoise} where all other
parameters are known. We started at very large $t_w$ when the
system is relaxed and $T_{eff}=T$ for all frequencies. Inserting
the values in eq.\ref{Vnoise} and using the $S_V$ measured at
$t_w=1 day$ we find $T_{eff}\simeq 333K$, within error bars for
all frequencies (see Fig.~\ref{noise}b). At short $t_w$ data show
that $T_{eff}(f,t_w)\simeq T_f$ for $f$ larger than a cutoff
frequency $f_o(t_w)$ which is a function of $t_w$. In contrast,
for $f<f_o(t_w)\ \ $ we find that $T_{eff}$ is:
$T_{eff}(f,t_w)\propto f^{-A(t_w)}$, with $A(t_w)\simeq 1$. This
frequency dependence of $T_{eff}(f,t_w)$ is quite well
approximated by

\begin{equation} T_{eff}(f,t_w)= T_f \ [ \ 1 \ + \ ( {f \over
f_o(t_w)})^{-A(t_w)} \ ]
 \label{fitTeff}
\end{equation}

where $A(t_w)$ and $f_o(t_w)$ are the fitting parameters. We find
that $1<A(t_w)<1.2$ for all the data set. Furthermore for $t_w
\geq 250$, it is enough to keep $A(t_w)=1.2$ to fit the data
within error bars. For $t_w <250s$ we fixed $A(t_w)=1$. Thus the
only free parameter in eq.\ref{fitTeff} is $f_o(t_w)$. The
continuous lines in Fig.~\ref{noise}(a) are the best fits of $S_V$
found inserting eq.\ref{fitTeff} in eq.\ref{Vnoise}.

In Fig.~\ref{noise}(b) we plot the estimated $T_{eff}(f,t_w)$ as a
function of frequency at different $t_w$. We see that just after
the quench $T_{eff}(f,t_w)$ is much larger than $T_f$ in all the
frequency interval. High frequencies rapidly decay towards the FDT
prediction whereas at the smallest frequencies $T_{eff}\simeq
10^5K$. Moreover we notice that low frequencies decay more slowly
than high frequencies and that the evolution of $T_{eff}(f,t_w)$
towards the equilibrium value is very slow. From the data of
Fig.~\ref{noise}(b) and eq.\ref{fitTeff}, it is easy to see that
$T_{eff}(f,t_w)$ can be superposed onto a master curve by plotting
them as a function of $f/f_o(t_w)$. The function $f_o(t_w)$ is a
decreasing function of $t_w$, but the dependence is not a simple
one, as it can be seen in the inset of Fig.~\ref{noise}(b). The
continuous straight line is not fit, it represents
$f_o(t_w)\propto 1/t_w$ which seems a reasonable approximation for
these data for $t>1000s$. For $t_w > 10^4 s$ we find the
$f_o<1Hz$. Thus we cannot follow the evolution of $T_{eff}$
anymore because the contribution of the experimental noise on
$S_V$ is too big, as it is shown in Fig.~\ref{noise}(b) by the
increasing of the error bars for $t_w=1 \ day$ and $f<0.1 Hz$.

\begin{figure}[ht!]
\centerline{\hspace{1cm} \bf (a) \hspace{6cm} (b) }
\begin{center}
\includegraphics[width=7cm]{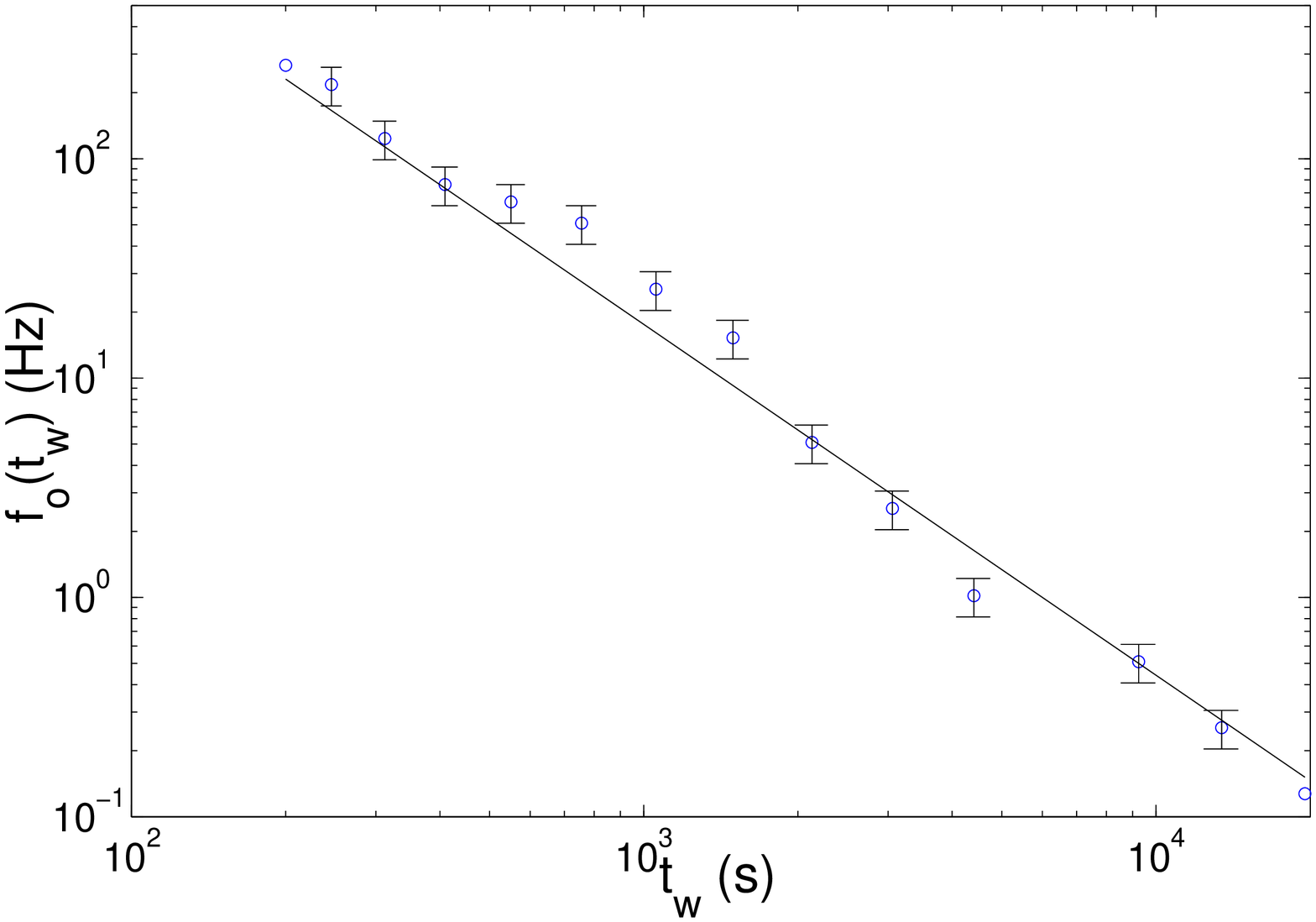}
\includegraphics[width=7cm]{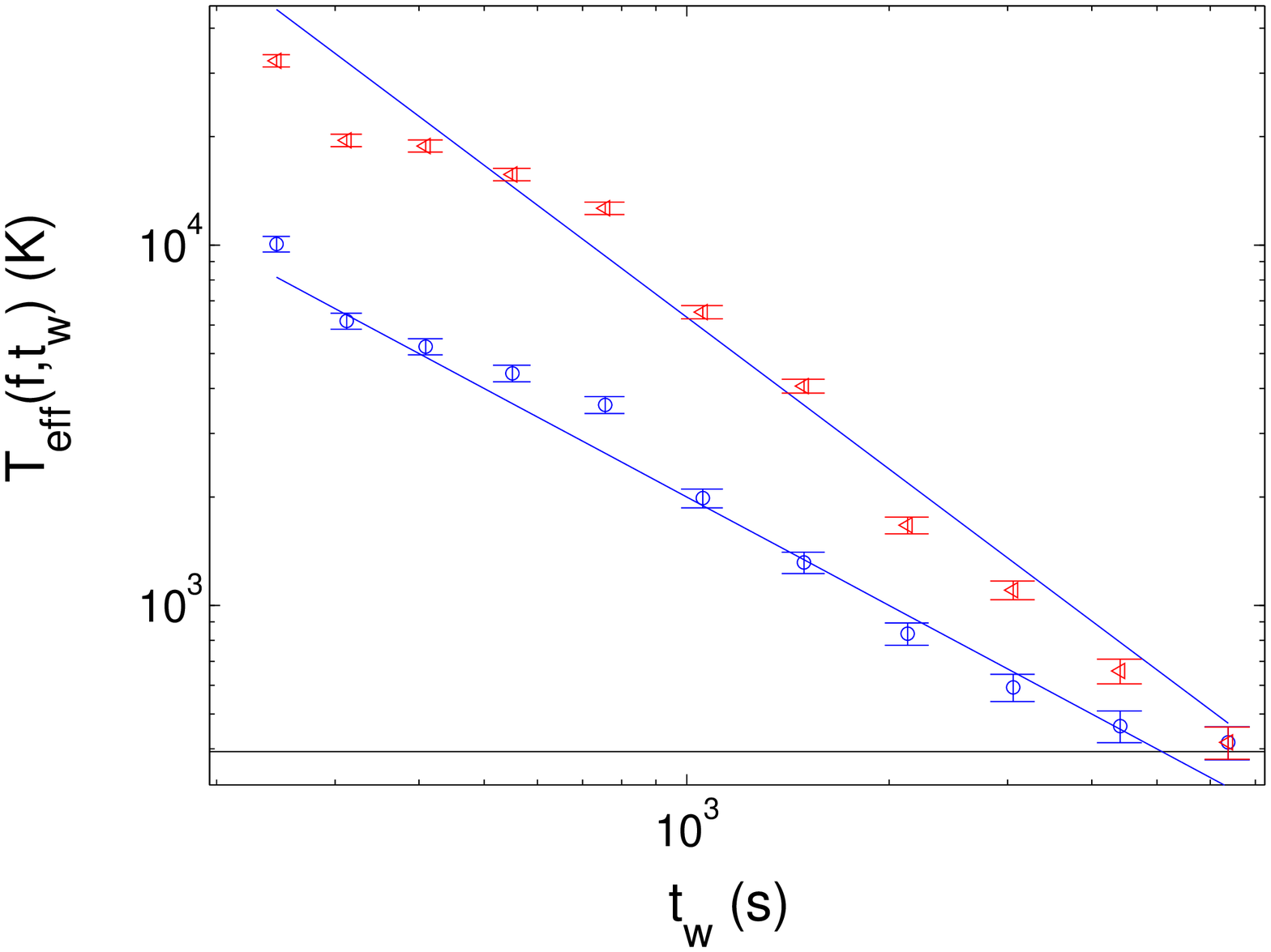}
\end{center}
\caption{{\bf $f_o$ and $T_{eff}$ as a function of $t_w$ at
$T_f=0.93T_g$ after a fast quench}. (a) $f_o$ defined in
eq.\ref{fitTeff} as a function of $t_w$ (b) Evolution of $T_{eff}$
at two different frequencies $(\circ) \ 7Hz$ and $(\triangleleft)
\ 2Hz$ } \label{Teffvf093Tg}
\end{figure}

We do not show the same data analysis for the other working
temperature after a fast quench, because the same scenario appears
in the range $0.79T_g<T<0.93T_g$, where the low frequency
dielectric properties are almost temperature independent (see
Fig.~\ref{hist}(b)). The only important difference to mention here
is that aging becomes faster and more pronounced as the
temperature increases. At $T_f=0.93T_g$, the losses of the
capacitor change of about $50\%$ in about $3h$, but all the
spectral analysis performed after a fast quench gives the same
evolution. We can just notice that $T_{eff}$ for $T=0.93 T_g$ is
higher than that at $T=0.79T_g$. At $T=0.93 T_g$, $T_{eff}$ is
well fitted by eq.\ref{fitTeff}. It is enough to keep $A(t_w)=1$
for all $t_w$ and $f_o(t_w)\sim 1/t_w^{1.5}$, see
fig.\ref{Teffvf093Tg}a). We notice that at $0.93T_g$ the power law
behaviour is well established, whereas it was more doubtful at
$0.73T_g$.  The dependence of $T_{eff}$ as a function of $t_w$ is
plotted in fig.\ref{Teffvf093Tg} for two values of $f$ and has
also a power law dependence on $t_w$.

For $T>0.93T_g$ fast quenches cannot be performed for the
technical reasons mentioned at the beginning of section 3). The
results are indeed quite different. Thus we will not consider, for
the moment, the measurement at $T_f=0.98T_g$ and we will mainly
focus on the measurements done in the range $0.79T_g<T<0.93T_g$
with fast quenches. For these measurements the spectral analysis
on the noise signal indicates that Nyquist relation (FDT) is
strongly violated for a long time after the quench. The question
is now to understand the reasons of this violation.

\subsection{ Statistical analysis of the noise}
\begin{figure}[ht]
\centerline{\hspace{1cm} \bf (a) \hspace{8cm} (b) }
\begin{center}
 \includegraphics[width=8cm]{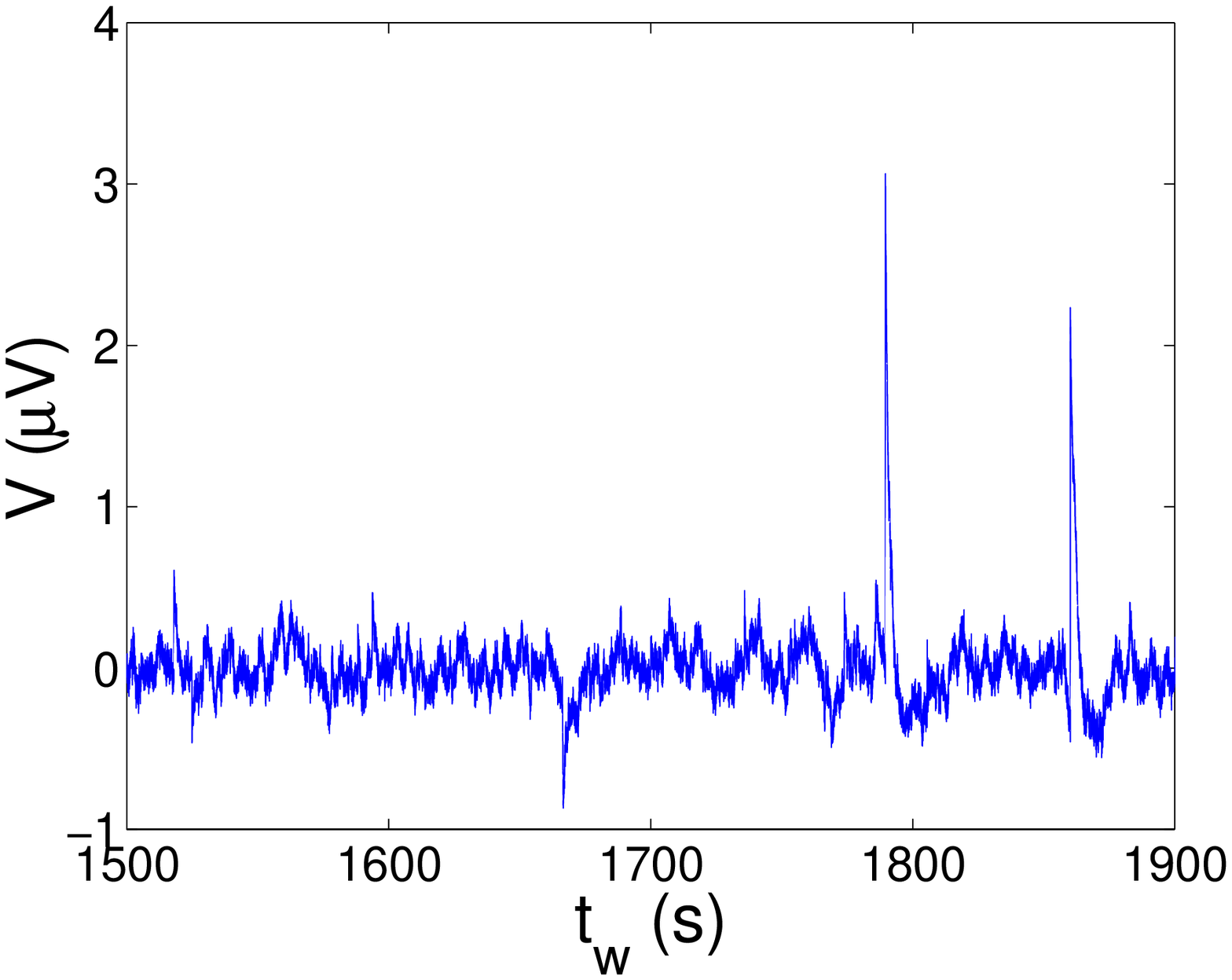}
 \hspace{1mm}
 \includegraphics[width=8cm]{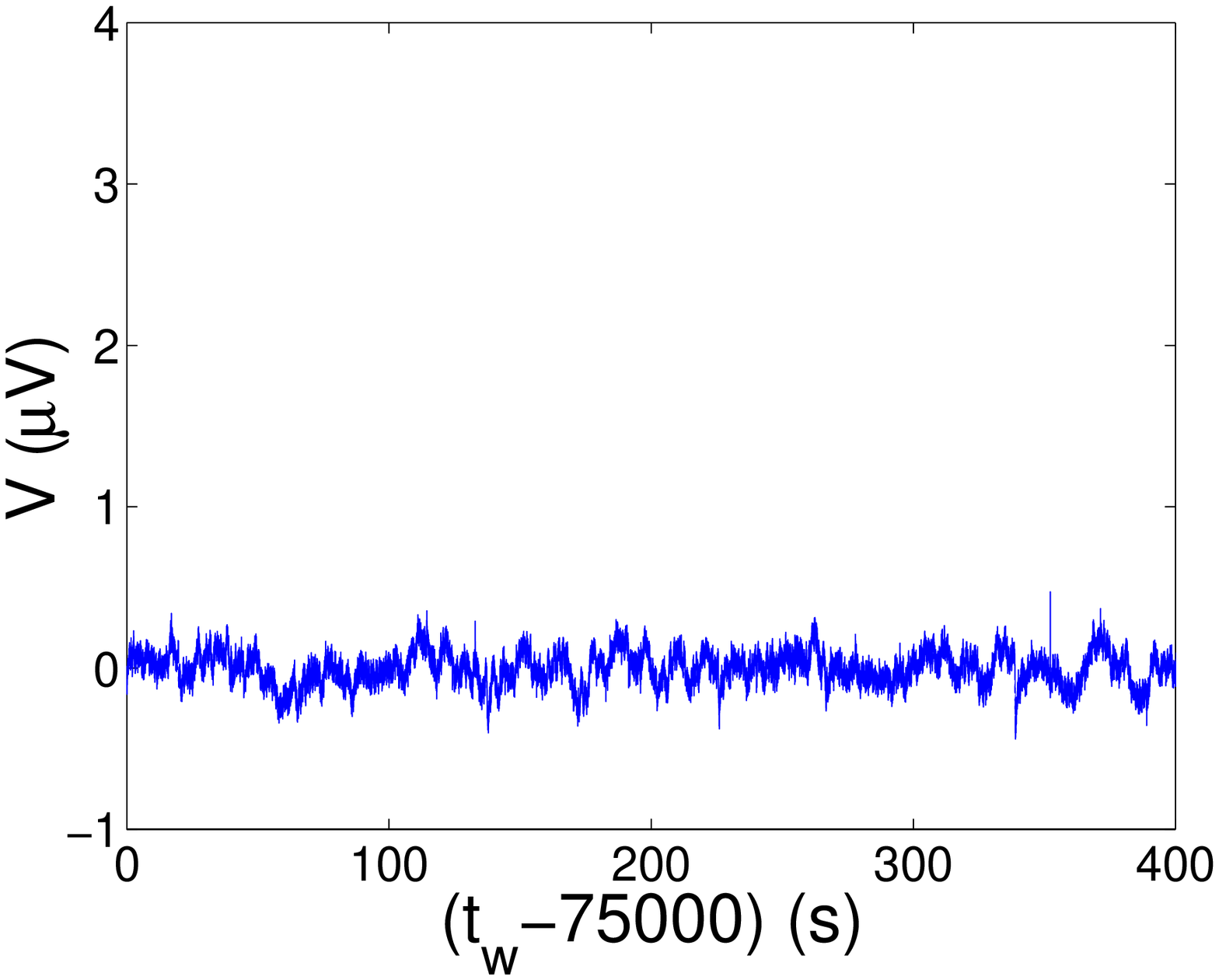}
 \end{center}
\caption{{\bf Voltage noise signal in polycarbonate after a fast
quench} Typical noise signal of polycarbonate measured at
$T_f=333K$ for  $1500s<t_w<1900s$ (a) and $t_w>75000s $ (b)}
 \label{signalpolyca}
\end{figure}

 In order to
understand the origin of such large deviations in our experiment
we have analyzed the noise signal. We find that the signal is
characterized by large intermittent events which produce low
frequency spectra proportional to $f^{-\alpha}$ with $\alpha
\simeq 2$. Two typical signals recorded at $T_f=0.79T_g$ for
$1500\,s<t_w<1900\,s$ and $t_w>75000\,s$ are plotted in
Fig.~\ref{signalpolyca}. We clearly see that in the signal
recorded for $1500\,s<t_w<1900\,s$ there are very large bursts
which are on the origin of the frequency spectra discussed in the
previous section.  In contrast in the signal which was recorded at
$t_w>75000\,s$, when FDT is not violated, the bursts totally
disappear (Fig.~\ref{signalpolyca}b).

\begin{figure}[!ht]
\centerline{\hspace{1cm} \bf (a) \hspace{8cm} (b) }
\begin{center}
\includegraphics[width=8cm]{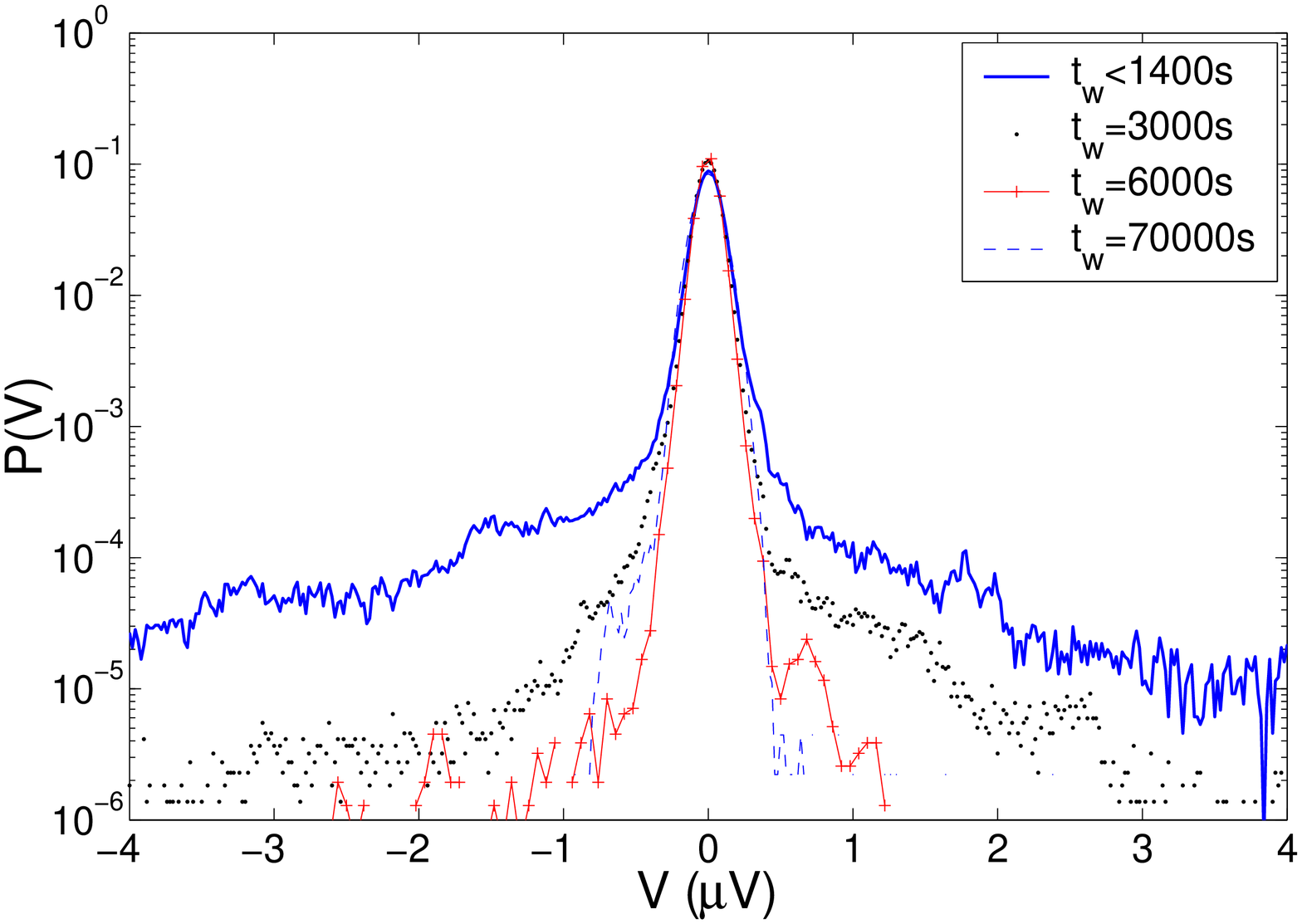}
\hspace{1mm}
\includegraphics[width=8cm]{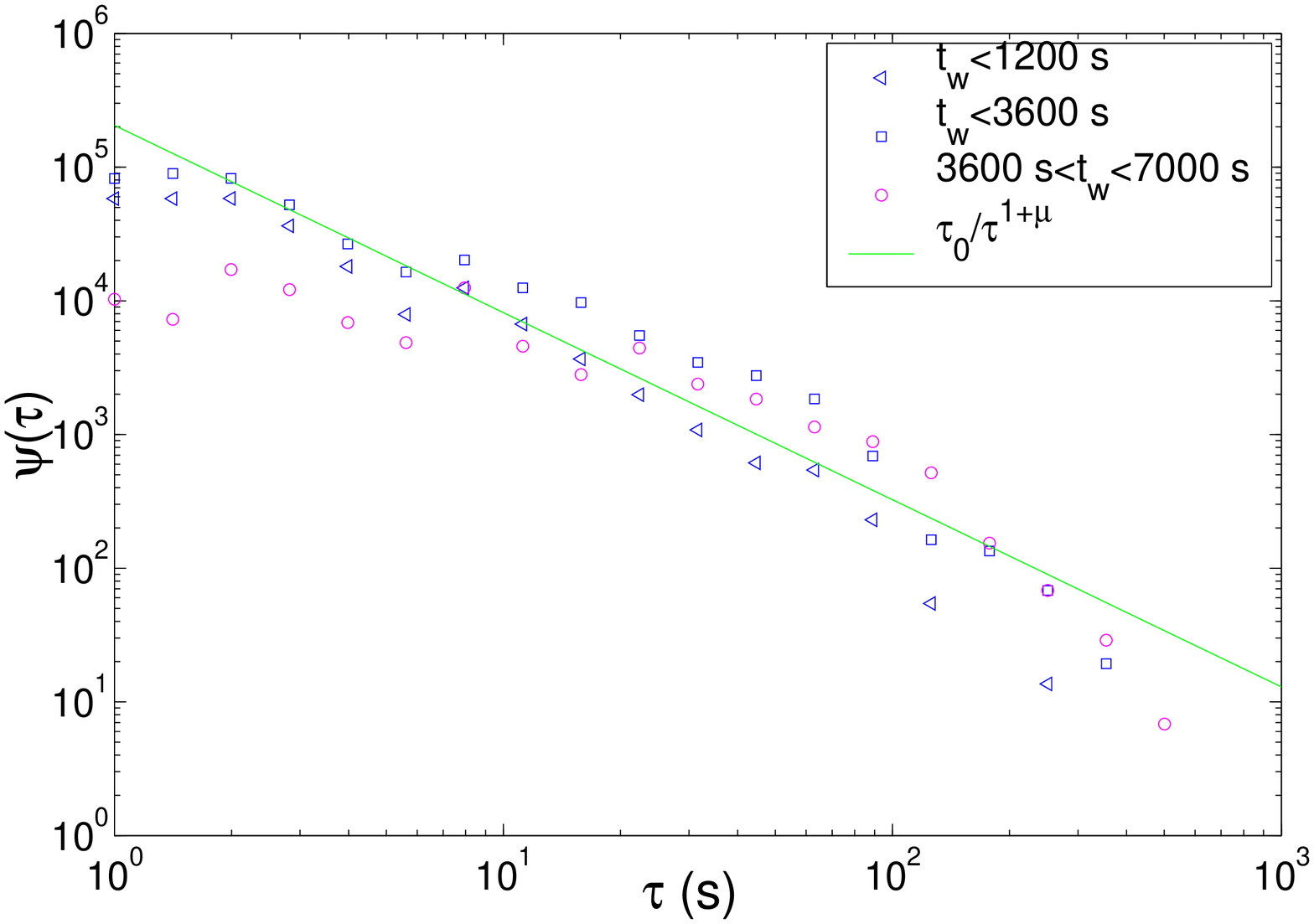}
\end{center}
\caption{{\bf PDF of voltage noise in polycarbonate after a fast
quench at $T_f=0.79T_g$.} (a) The large tails of the PDF at early
$t_w$ are a signature of strong intermittency.  (b) Histogram of
time interval $\tau$ between two successive pulses: $\Psi(\tau,
t_w)$. At early $t_w$, $\Psi(\tau, t_w)$ is power law
distributed.}
 \label{PDFpolyca}
\end{figure}

The probability density function (PDF) of these signals is shown
in Fig.~\ref{PDFpolyca} (a). We clearly see that the PDF, measured
at small $t_w$, has very high tails which becomes smaller and
smaller at large $t_w$. Finally the Gaussian   profile   is
recovered after $24h$. The PDF are very symmetric in their
gaussian parts, {\it i.e.} 3 standard deviations. The tails of the
PDF are exponential and are   decreasing functions   of $t_w$.

The time interval $\tau$ between two successive pulses is power
law distributed. In order to study   this   distribution
$\Psi(\tau, t_w)$ of $\tau$, we have first selected the signal
fluctations with amplitude larger than a fixed threshold, which
  has been chosen between 3 and 4 standard deviations of the
equilibrium noise, \textit{i.e.} the noise predicted by the FDT.
We have then measured the time intervals $\tau$ between two
successive large fluctuations. The histogram $\Psi(\tau, t_w)$
computed for $t_w<20min$ and for $20min<t_w<3h$ is plotted in
Fig.~\ref{PDFpolyca}(b).   We clearly see that $\Psi(\tau, t_w)$
is a power law, specifically
$\Psi(\tau)\propto\frac{1}{\tau^{1+\mu}}$ with $\mu\simeq 0.4\pm
0.1$. This   result   agree with one of the hypothesis of the trap
model\cite{Bouchaud}-\cite{trap}, which presents non-trivial
violation of FDT associated to an intermittent dynamics. In the
trap model $\tau$ is a power-law-distributed quantity with an
exponent 1+$\mu$ that, in the glass phase, is smaller than 2.
However, there are important differences between the dynamics of
our system and that of the trap model. Indeed in this model one
finds short and large $\tau$ for any $t_w$ which is in contrast
with our system because the probability of finding short $\tau$
seems to decrease as a function of $t_w$. But this effect could be
a consequence of the imposed threshold. It seems that there is no
correlation between the $\tau$ and the amplitude of the associated
bursts. Finally, the maximum distance $\tau_{max}$ between two
successive pulses grows as a function of $t_w$ logarithmically,
that is $\tau_{max}=[10+152log(t_w/300)]s$ for $t_w>300s$. This
slow relaxation of the number of events per unit of time shows
that the intermittency is related to aging.

\begin{figure}[!ht]
\centerline{\hspace{1cm} \bf (a) \hspace{8cm} (b) }
\begin{center}
\includegraphics[width=8cm]{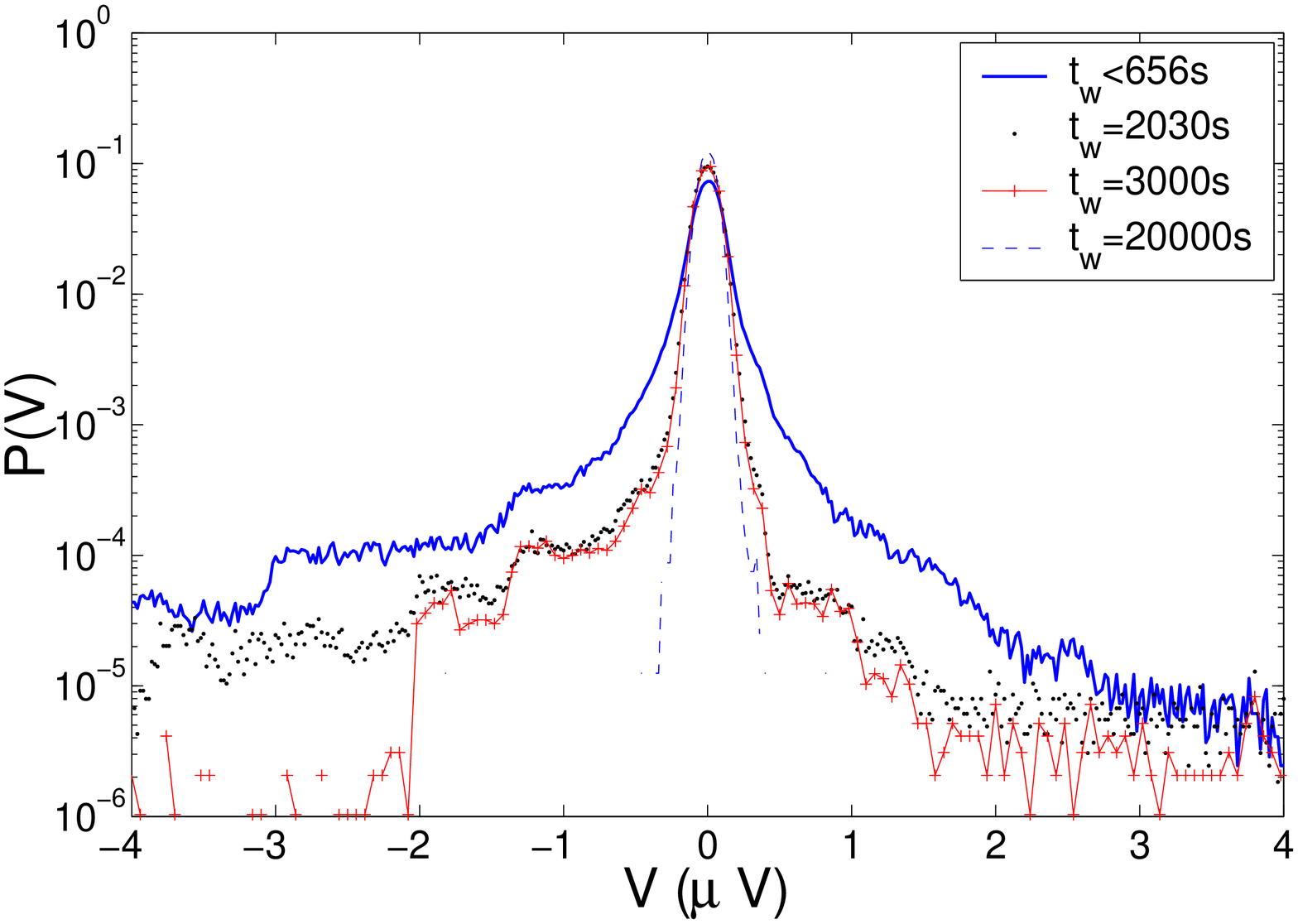}
\hspace{1mm}
\includegraphics[width=8cm]{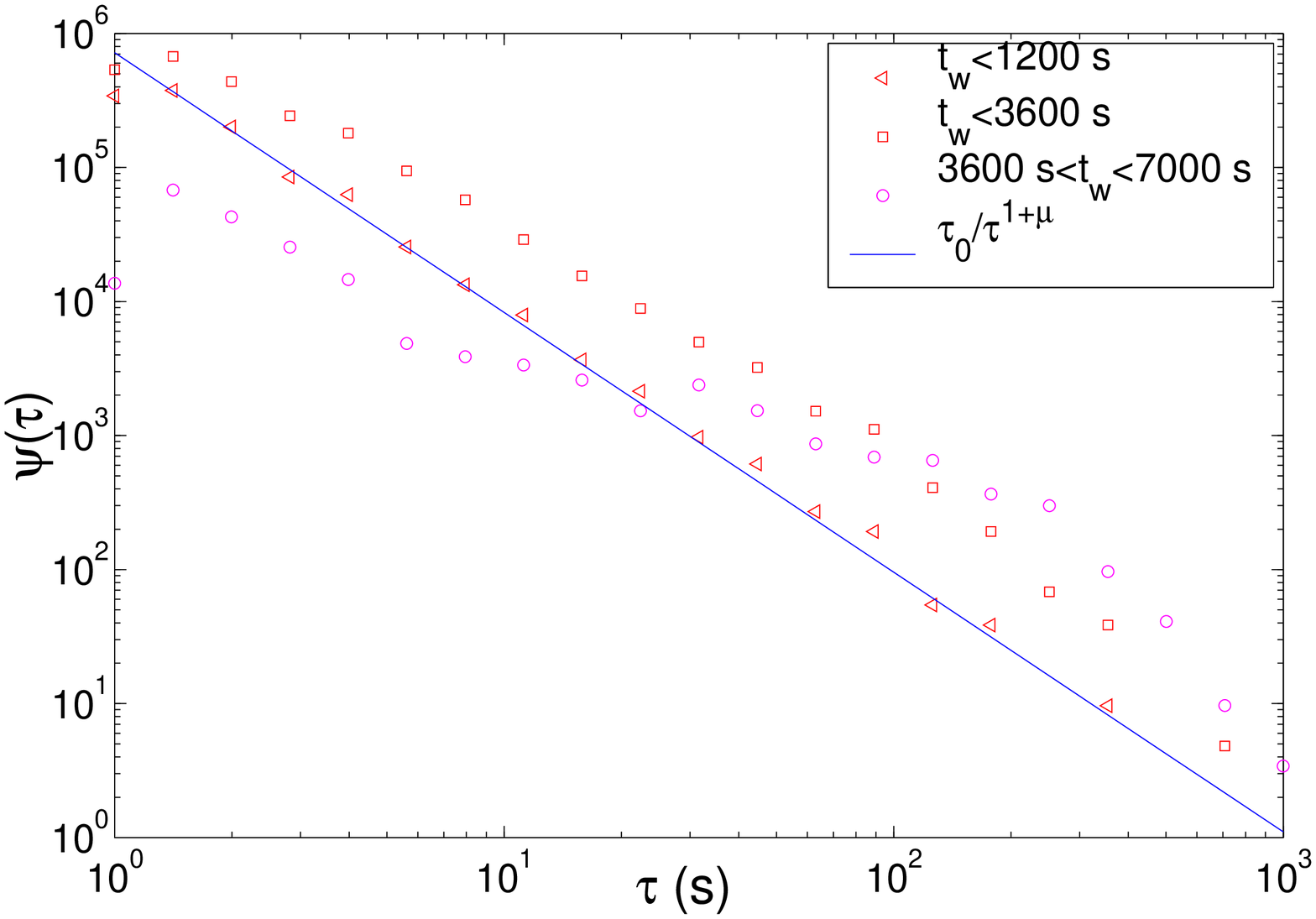}
\end{center}
\caption{{\bf PDF of voltage noise in polycarbonate after a fast
quench at $0.93 T_g$}(a)  PDF of the noise signal of polycarbonate
measured at various $t_w$.  (b) Histograms $\Psi(\tau, t_w)$,
following at early $t_w$ a $1/\tau^{1+\mu}$ law with $\mu=0.9 \pm
0.1$.  } \label{PDF120}
\end{figure}

The same behaviour is observed at $T_f=0.93T_g$ after a fast
quench. The PDF of the signals measured at $T_f=0.93T_g$ are shown
in Fig.~\ref{PDF120}(a). The behaviour is the same except for the
relaxation rate towards the Gaussian distribution which is faster
in this case, because the aging effects are larger at this
temperature. From these measurements one concludes that after a
fast quench the electrical thermal noise is strongly intermittent
and non-Gaussian. The number of intermittent events increases with
the temperature : for $T_f=0.93 T_g$, $T_{eff}$ is higher than for
$T_f=0.79 T_g$ and PDF tails are more important.
The histograms $\Psi(\tau, t_w)$ are shown Fig.~\ref{PDF120} (b).
The behaviour is the same with   $\mu=0.9 \pm 0.1$.  By comparing
$\Psi(\tau, t_w)$ for short $\tau$ and short $t_w$ there are more
events at $0.93 T_g$ (Fig.~\ref{PDF120} (b)) than at $0.79 T_g$
(Fig.~\ref{PDFpolyca} (b)). This is consistent with activation
processes for the aging dynamics. Indeed the probability of
jumping from a potential well to another increases with
temperature. Thus one expects to find more events at high
temperature than at low temperature.

\subsection{Influence of the quench speed.} 
\begin{figure}[!ht]
\centerline{\hspace{1cm} \bf (a) \hspace{8cm} (b) }
\begin{center}
\includegraphics[width=8cm]{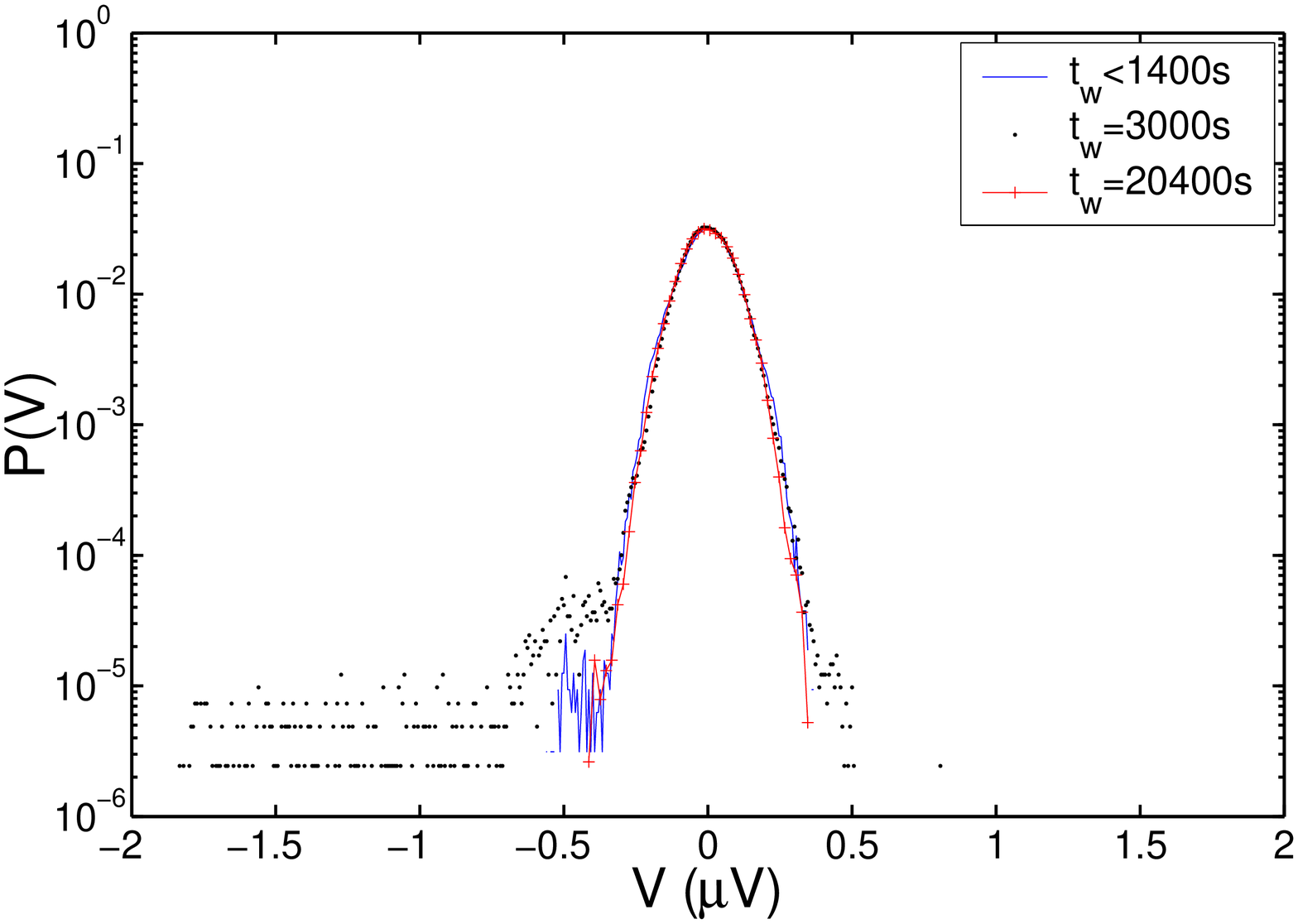}
\hspace{1mm}
\includegraphics[width=8cm]{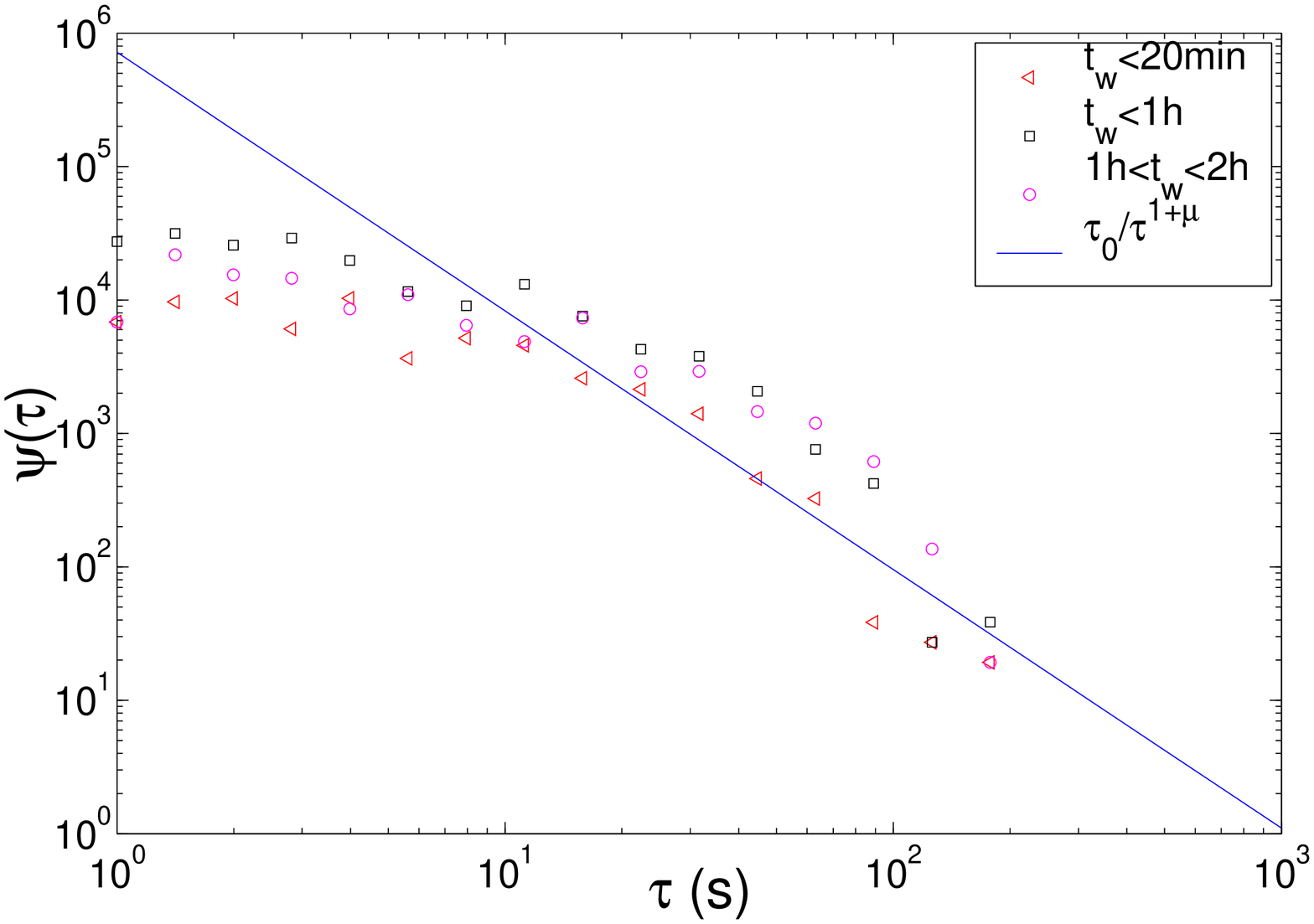}
\end{center}
\caption{{\bf PDF of voltage noise in polycarbonate after a slow
quench at $T_f=0.93T_g$.} (a) No intermittency is visible after a
slow quench at $3.6 K/min$. (b) Histograms $\Psi(\tau, t_w)$ after
a slow quench.   The line corresponds to the $\tau_0/\tau^{1+\mu}$
fit of Fig.~\ref{PDF120}(b), where $\mu = 0.93$.   }
\label{PDF120lent}
\end{figure}

The intermittent
behaviour  described in the previous sections depends on the
quench speed. In Fig.~\ref{PDF120lent}(a) we plot the PDF of the
signals measured after a slow quench ($3.6\,K/min$)   at $T_f =
0.93 T_g$.   We clearly see that the PDF are very   different:
intermittency has almost disappeared.   The comparison between the
fast quench and the slow quench merits a special comment. During
the fast quench $T_f=0.93T_g$ is reached in about $100\,s$ after
the passage of $T$ at $T_g$. For the slow quench this time is
about $1000\,s$. Therefore one may wonder whether after $1000s$ of
the fast quench one recovers the same dynamics of the slow quench.
By comparing the PDF of Fig.~\ref{PDF120}(a) with those of
Fig.~\ref{PDF120lent}(a) we clearly see that this is not the case.
Furthermore, by comparing the histograms of Fig. \ref{PDF120}(b)
with those of Fig.\ref{PDF120lent}(b), we clearly see that there
are less events separated by short $\tau$ for the slow quench.
Therefore one deduces that the polymer is actually following a
completely different dynamics after a fast or a slow quench
\cite{BertinKovacs,Sciortino}. This is a very important
observation that can be related to well known effects of response
function aging.  The famous Kovacs effect is an
example\cite{Kovacs} where depending on the cooling rate the
isothermal compressibility presents a completely different time
evolution.

\begin{figure}[ht!]
\begin{center}
\includegraphics[width=8cm]{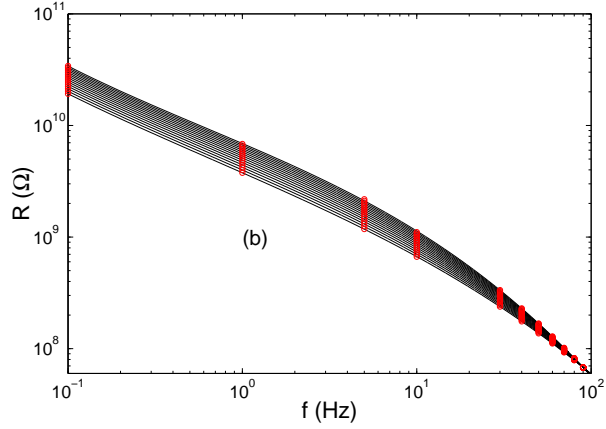}
\end{center}
\caption{{\bf Capacitance losses at $T_f=0.98T_g$ after a slow
quench}.  Resistance as a function of frequency for different
$t_w$ from $t=100s$ (lower curve) to $t_w=14400s$.} \label{R098Tg}
\end{figure}

\begin{figure}[ht!]
\begin{center}
\includegraphics[width=8cm]{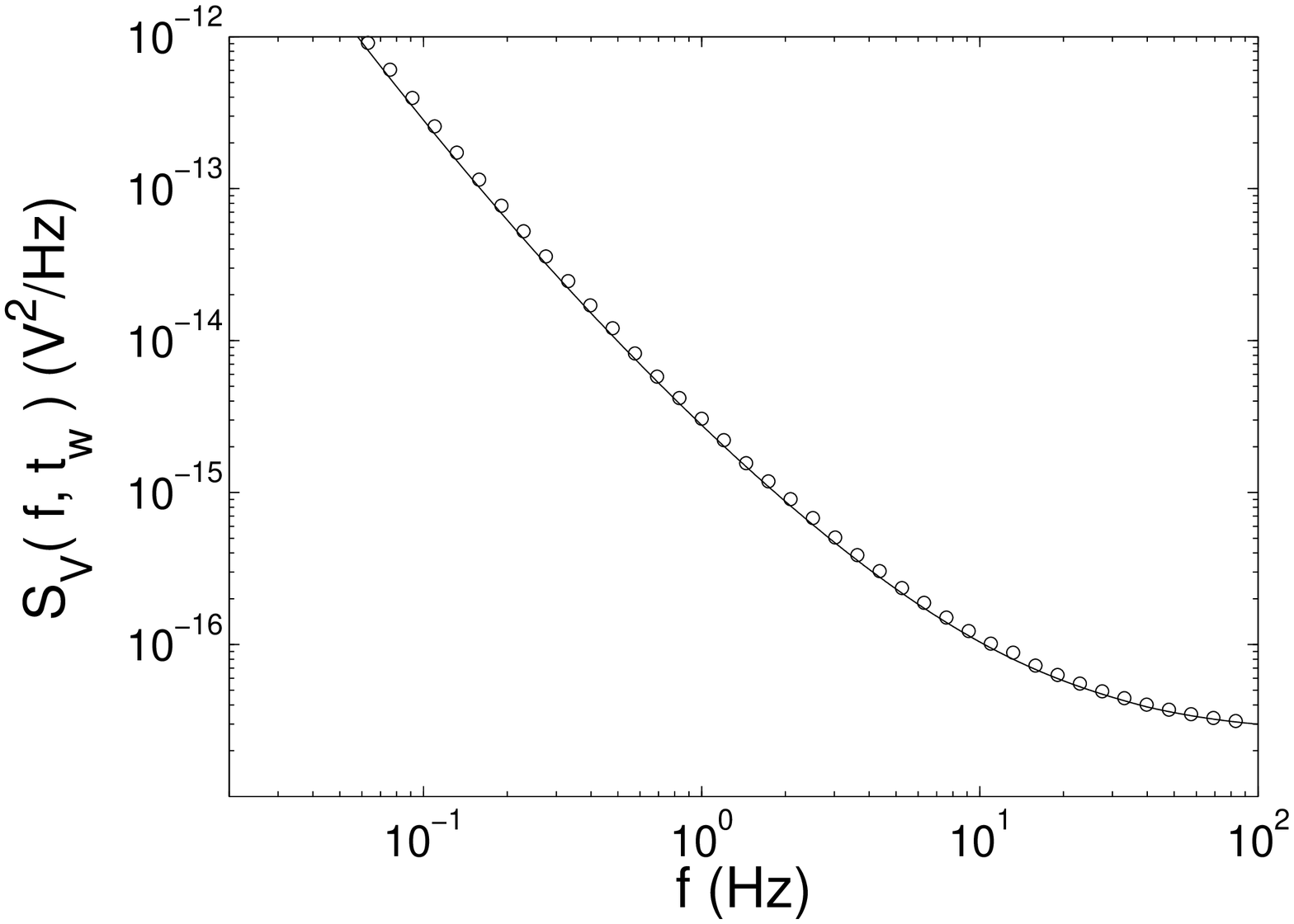}
\hspace{1mm}
\includegraphics[width=8cm]{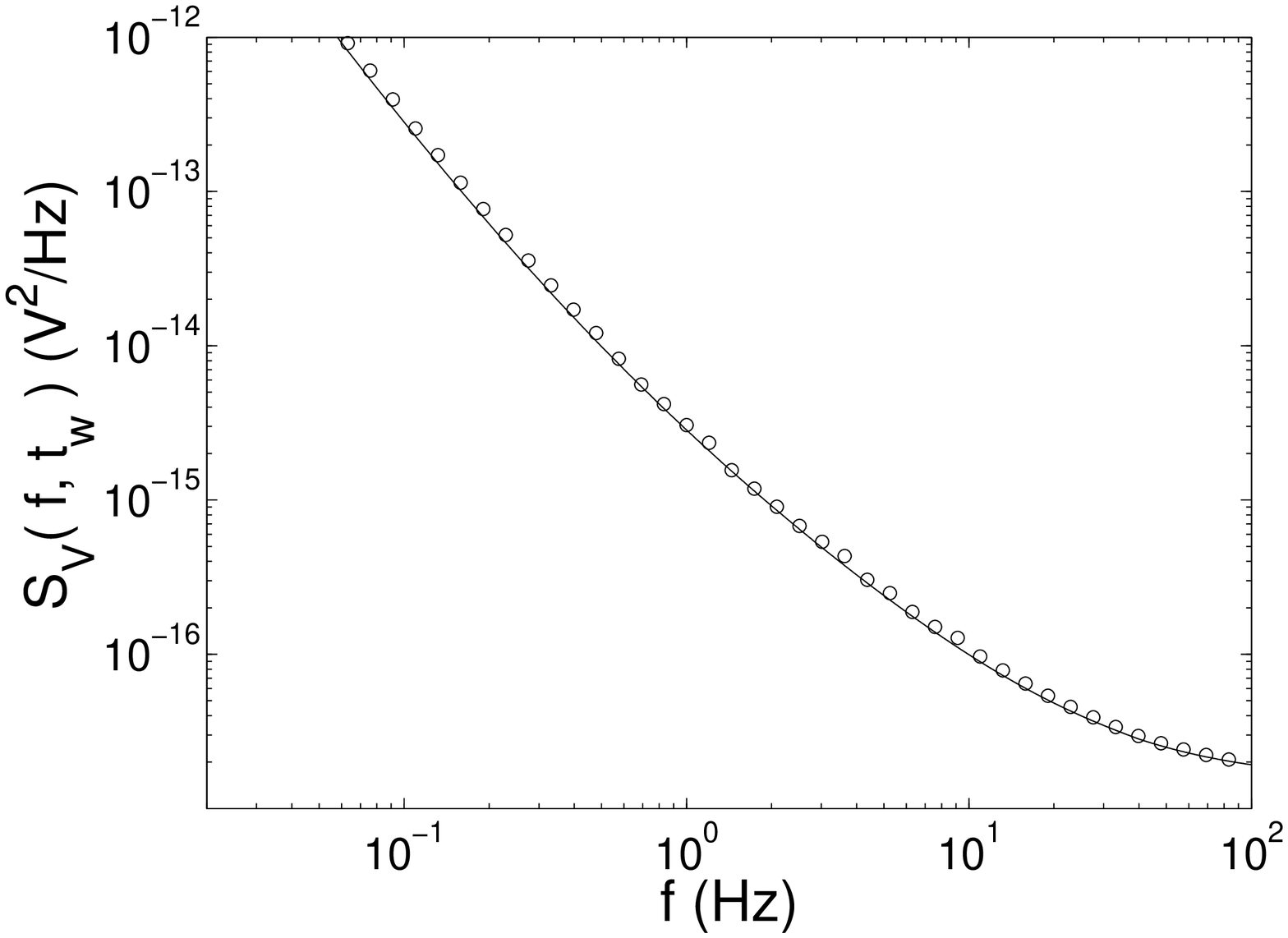}

\end{center}
\caption{{\bf Power spectral density of the capacitance noise  at
$T_f=0.98T_g$ after a slow quench}. $S_V(f,t_w)$ as a function of
$f$ for two different time: (a) $t_w=200s$, (b) $t_w=7200s$.
  Circles stand for measurement points, whereas the continuous line is the FDT prediction.  }
\label{noise098Tg}
\end{figure}

\begin{figure}[!]
\begin{center}
\includegraphics[width=8cm]{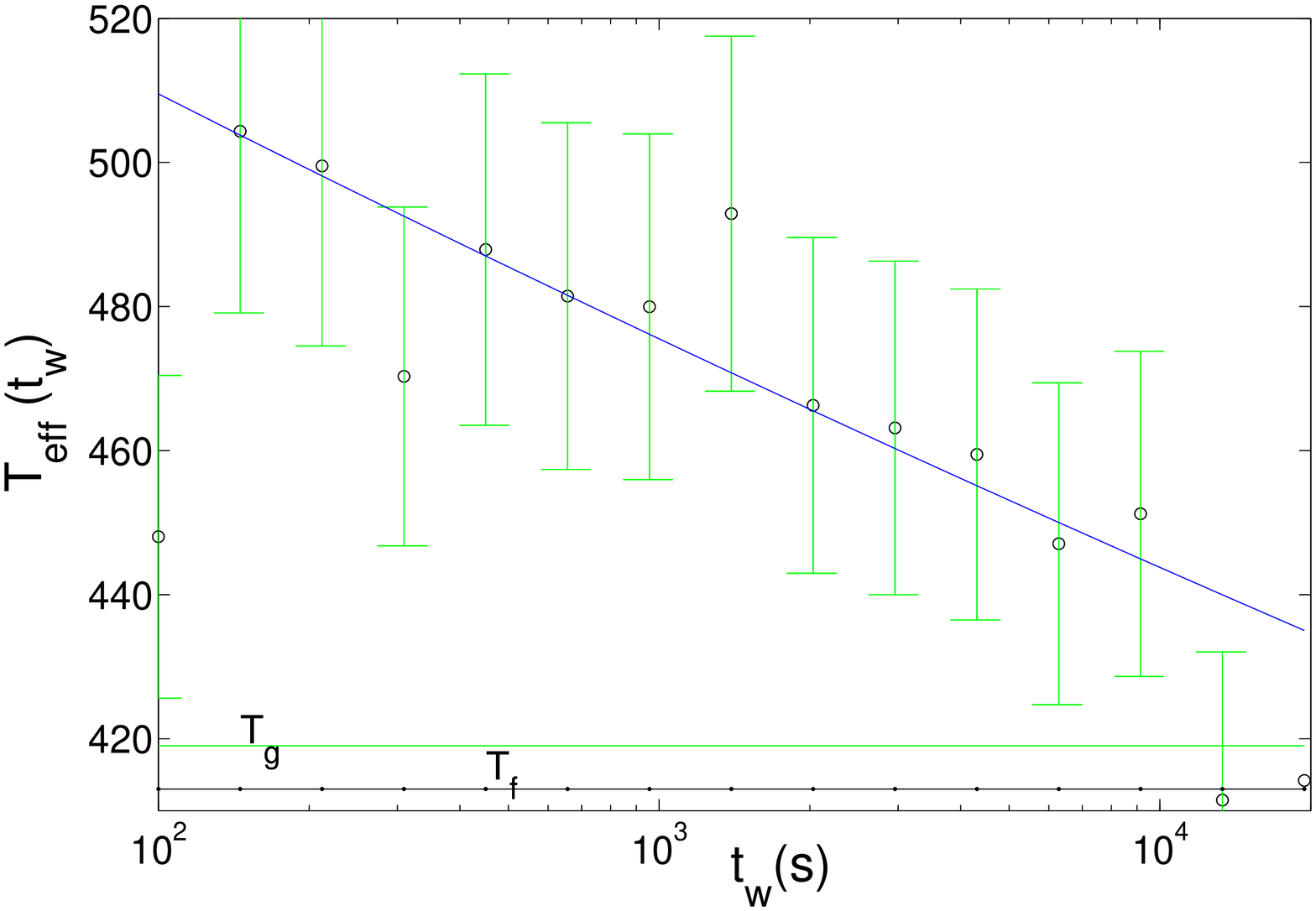}
\end{center}
\caption{{\bf $T_{eff}$ as a function of time at $T_f=0.98T_g$
after a slow quench}. $T_{eff}$ averaged in the frequency band
$[1Hz-10Hz]$. It has been computed from the spectra $S_V(f,t_w)$
and the measured $(R,C)$ using eq.\ref{Vnoise} }.
\label{Teff098Tg}
\end{figure}

\begin{figure}[ht!]
\centerline{\hspace{1cm} \bf (a) \hspace{8cm} (b) }
\begin{center}
\includegraphics[width=8cm]{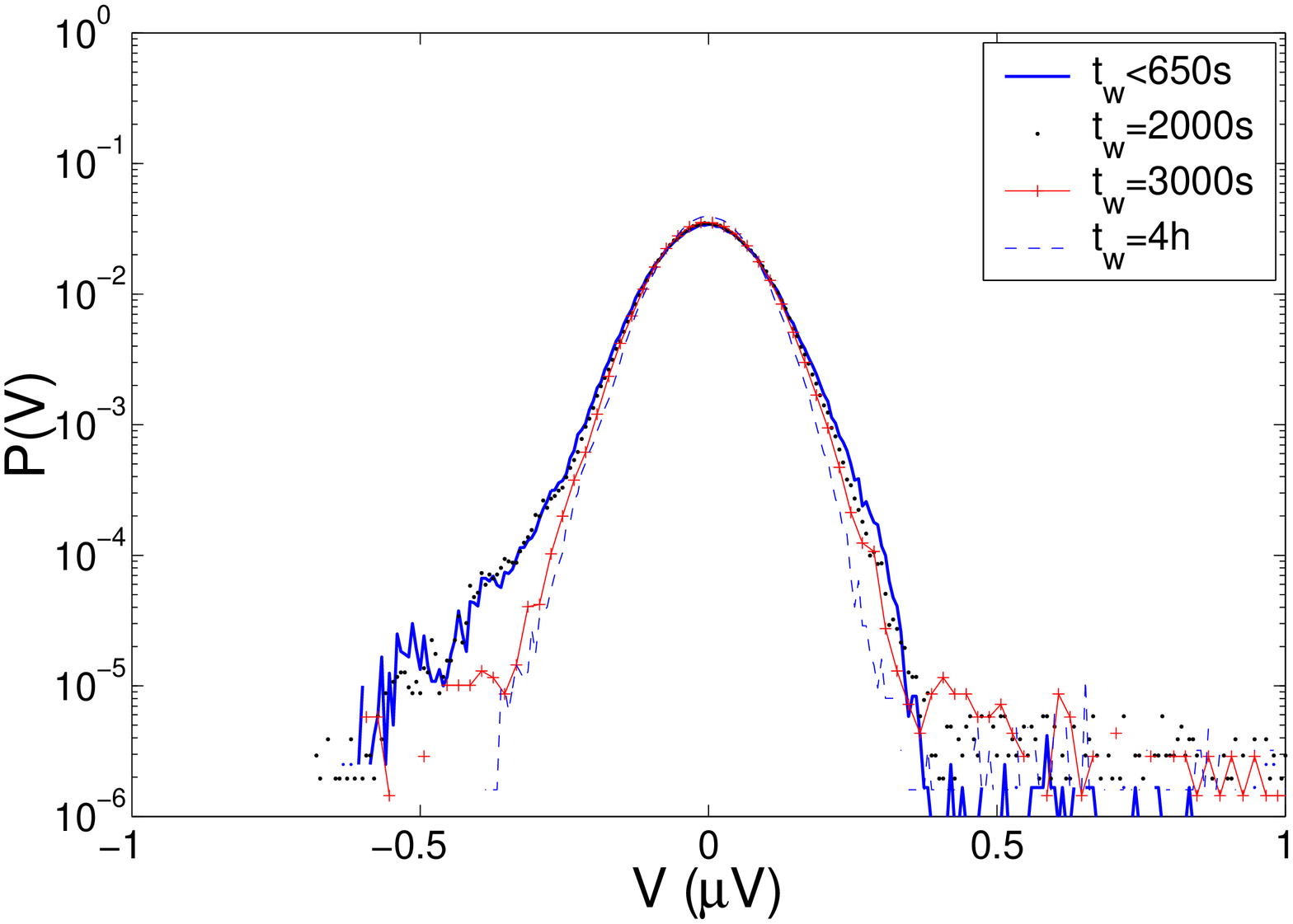}
\hspace{1mm}
\includegraphics[width=8cm]{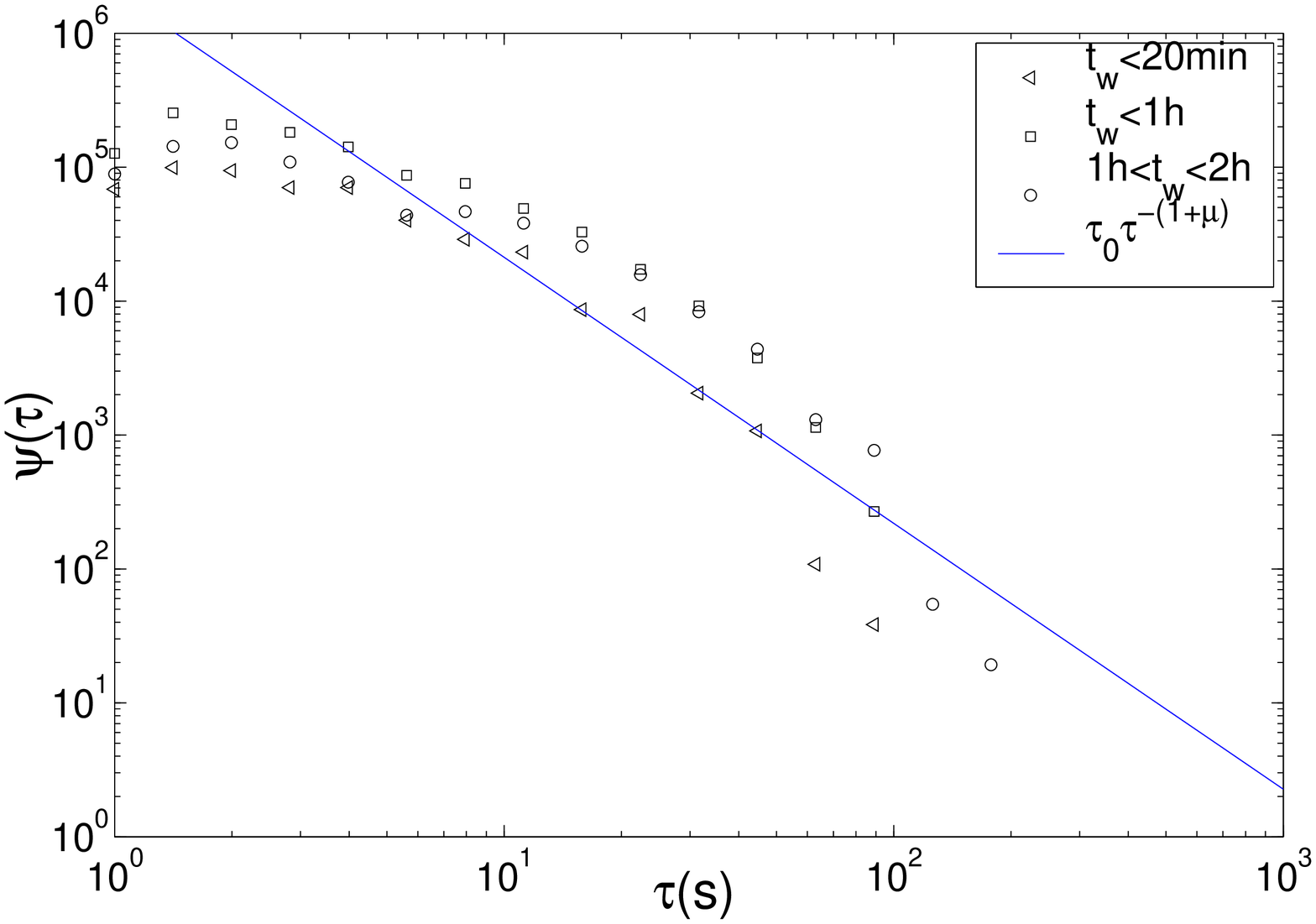}
\end{center}
\caption{{\bf PDF of the signal at $T_f=0.98T_g$ after a slow
quench}. (a) PDF of the signal. No intermittency is visible. (b)
Histograms $\Psi(\tau, t_w)$ after a slow quench.   The continuous
line represent the $1/\tau^{1+\mu}$ law where we chose
$\mu=T_f/T_g$  to compare with the theoretical estimation of the
trap model.   } \label{PDF098Tg}
\end{figure}

\subsection{$T_{eff}$ after a slow quench}

In the previous section we have shown that the intermittent aging
dynamics is strongly influenced  by the cooling rate. We discuss
in this section the behaviour of the effective temperature after a
slow quench. We use for this purpose the measurement at
$T_{f}=0.98T_g$. The time evolution of the response function is
much larger  at   this temperature   than at $T_f=0.79T_g$ as it
can be seen in fig.\ref{R098Tg}. It is about $50\%$ at the small
frequencies, therefore it has to be kept into account in the
evaluation of FDT. The spectrum of the capacitance noise measured
at $0.98T_g$ is plotted for two different times in
fig.\ref{noise098Tg}. The continuous lines represents the FDT
predictions computed using the measured response function reported
in fig.\ref{R098Tg}. We clearly see that the experimental points
are very close to the FDT predictions, thus the violation of FDT,
if it exists,  is very small. To check this point, we have
computed $T_{eff}$ in the range $[1Hz-10Hz]$, which is plotted as
a function of time in fig.\ref{Teff098Tg}. Although the error bars
are rather large, we clearly see that $T_{eff}$ decreases
logarithmically as a function of time. We also notice that the
maximum violation at short times is about $25\%$ which  is  much
smaller than that measured at smaller $T_f$  after  a fast quench.
The PDF of the noise signal at $0.98T_g$ are plotted in
fig.\ref{PDF098Tg}(a) and they  do not show very large tails as in
the case of the intermittent dynamics. The statistics of the time
intervals $\tau$ between two large events does not present  any
power law   either (see fig.\ref{PDF098Tg}(b)).   Thus the signal
statistics  look much more similar to those measured at $0.93T_g$
after a slow quench than   to   the intermittent ones. This
comparison shows that independently of the final temperature the
intermittent behaviour is induced by the fast quench and that the
FDT violation is cooling rate dependent.

\section{ Mechanical measurements on a Polycarbonate cantilever}

In the previous section we have studied the properties of
dielectrical thermal noise during the aging of polycarbonate. In
this section we want  to check if the thermal noise features are
independent of the observable. As a second observable, we  have
chosen to measure the thermally excited vibrations of a cantilever
made of polycarbonate.

\subsection{ FDT in a mechanical oscillator}
 The physical object of our
interest is a small plate with one end clamped and the other free,
i.e. a cantilever. The plate is of length $l$, width $a$,
thickness $b$, mass $m_{\mathrm{Polyc}}$. On the free end of the
cantilever a small golden mirror of mass $m_{\mathrm{mirror}}$ is
glued. As described in the next section, this mirror is used to
detect the amplitude $x_c$ of the transverse vibrations of the
cantilever free end. The motion of the cantilever free end can be
assimilated to that of a driven harmonic oscillator, which is
damped only by the viscoelasticity of the polymer. Consequently,
the equation of motion of the cantilever free end takes a simple
form in Fourier space:
\begin{equation}
    [- m \omega^2 + K(\omega) ] \hat{x}_c = \hat{F}_{\mathrm{ext}}
     \label{HO}
\end{equation}

\noindent where $\hat{x}_c$ is the Fourier transform of $x_c$, $m$
is the total effective mass of the plate plus the mirror, $K = K'
+ i K''$ is the complex elastic stiffness of the plate free end,
and $\hat{F}_{\mathrm{ext}}$ is the Fourier transform of the
external driving force. The complex $K(\om)$ takes into account
the viscoelastic nature of the cantilever. From the theory of
elasticity \cite{bib1} one obtains that, for low frequencies, an
excellent approximations for $m$ and $K$ are:
\begin{eqnarray}
    m & = & \frac{3}{(3.52)^2} \, m_{\mathrm{Polyc}} + m_{\mathrm{mirror}}, \\
    \textrm{and } K & = & {E a b^3 \over 4 l^3},
\end{eqnarray}

\noindent where $E = E' + i E''$ is the plate Young modulus.
Notice that if $m_{\mathrm{mirror}} = 0$, then one recovers the
smallest resonant frequency of the cantilever \cite{bib1}. For
Polycarbonate at room temperature, $E$ is such that $E' = 2.2
\times 10^9 \textrm{ Pa}$ and $E'' = 2 \times 10^7 \textrm{ Pa}$,
and its frequency dependence may be neglected in the range of
frequency of our interest, that is from 0.1 to 100 Hz \cite{bib3}.
Thus we neglect the frequency dependence of $K$ in this specific
example.

When $F_{\mathrm{ext}} = 0$, the amplitude of the thermal
vibrations of the cantilever  free end $x_T$ is linked to its
response function $\chi$ via the FDT \cite{bib2}:

\begin{equation}
    \langle \vert{ \hat{x}_T \vert }^2 \rangle = \frac{2 k_B T}{\omega} \, \mathrm{Im} \, \hat{\chi}, \label{FDT}
\end{equation}

\noindent where $\langle {\vert \hat{x}_T \vert }^2 \rangle$ is
the thermal fluctuations spectral density of $x_c$, $k_B$ the
Boltzmann constant and $T$ the temperature. From Eq.\,\ref{HO} one
obtains that the response function of the harmonic oscillator is
\begin{equation}
    \hat{\chi} = \frac{\hat{x}_c}{\hat{F}_{\mathrm{ext}}} =
    { 1 \over m \lbrack {\omega_0}^2 - \om^2 - i \,
    (\mathrm{sign}\, \omega) \, \gamma {\omega_0}^2 \rbrack}, \label{eq4}
\end{equation}

\noindent where ${\omega_0}^2 = K' / m$ and $\gamma = K'' / K'$.

Inserting Eq.\,\ref{eq4} into Eq.\,\ref{FDT}, one can compute the
thermal fluctuations spectral density of the Polycarbonate
cantilever for positive frequencies:
\begin{equation}
    \langle { \vert \hat{x}_T \vert }^2 \rangle = \frac{2 k_B T}{\om}
    \frac{\gamma {\omega_0}^2}{m \lbrack ({\omega_0}^2 -
     {\omega}^2)^2 + (\gamma {\omega_0}^2)^2 \rbrack}. \label{eq5}
\end{equation}

\noindent Notice that $\langle {\vert{ \hat{x}_T }\vert}^2 \rangle
\sim \om^{-1}$ for $\om \ll \om_0$, because the viscoelastic
damping $K''$ is constant in our frequency range. In the case of a
viscous damping (for example, a cantilever immersed in a viscous
fluid) $K'' = \alpha \, \omega$, where $\alpha$ is proportional to
the fluid viscosity and to a geometry dependent factor. Then the
thermal fluctuations spectral density of the cantilever free end,
in the case of viscous damping, is
\begin{equation}
    \langle {\vert \hat{x}_T \vert}^2 \rangle = \frac{2 k_B T \, \alpha}
    {m^2 \lbrack (\omega_0^2-\omega^2)^2 + (\frac{\alpha}{m} \omega)^2 \rbrack }, \label{eq5b}
\end{equation}

\noindent which is constant for $\omega \ll \omega_0$. Therefore,
the thermal fluctuations spectral density shape depends on
$K''(\omega)$. In the case of a viscoelastic damping
(Eq.\,\ref{eq5}),   the thermal noise increases   when $\omega$
goes to 0,   and with a suitable choice of the parameters the low
frequency spectral density of an aging polymer can be   measured
  using this method.

\begin{figure}[!tbp]
    \begin{center}
    \includegraphics[width=10cm, angle=0]{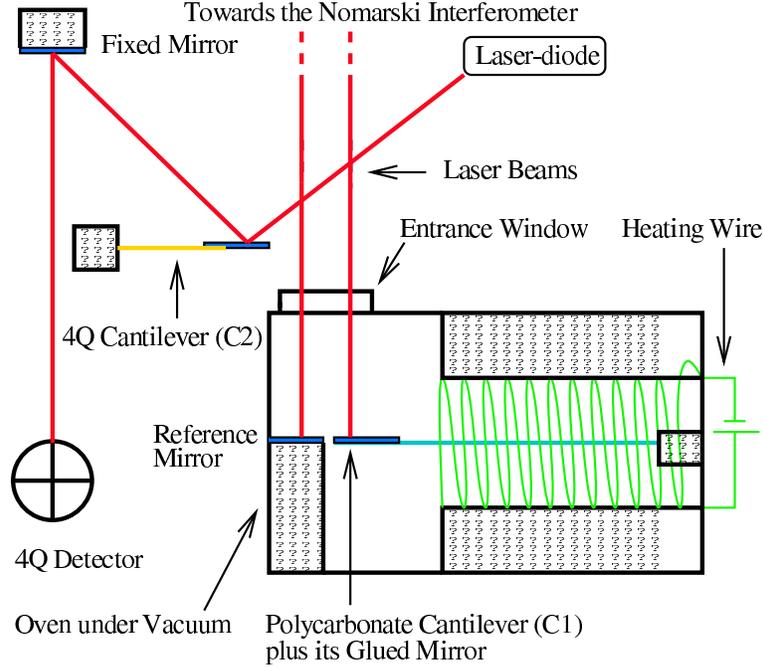}
    \caption{{\bf Experimental setup for measuring the mechanical noise in polycarbonate}
    The polycarbonate cantilever (C1) is inside an oven to control the
    temperature. Its displacement is measured by a very sensitive Nomarski Interferometer.
    The cantilever (C2) the laser diode, and the 4Q
    detector, are use for the noise reduction technique (see
    text).
    } \label{fig2}
    \end{center}
\end{figure}

However, the cantilever is also sensitive to the mechanical noise,
and the total displacement $x_c$ of the cantilever free end
actually reads $x_c = x_T + x_{\mathrm{acc}}$, where
$x_{\mathrm{acc}}$ is the displacement induced by the external
mechanical noise. Thus, it is important to compute the
signal-to-noise ratio (SNR) of our apparatus, which we define as
the ratio between the thermal fluctuations and the mechanical
noise spectral densities. All the details on the optimization of
the SNR can be found in ref.\cite{RSInoise}.

\subsection{ Experimental apparatus}

Let us estimate the amplitude of $\sqrt{ {\langle {\vert{
\hat{x}_T }\vert}^2 \rangle} }$ at $\nu = \om / 2\pi = 1 \textrm{
Hz}$ for the following choice of the parameters: $\gamma \simeq
10^{-2}$, $l \simeq 10 \textrm{ mm}$, $a \simeq 1 \textrm{ mm}$,
$b = 125\ \mu \mathrm{m}$ and $m_{\mathrm{mirror}} \lesssim
10^{-3} \textrm{ g}$. We find $\nu_0 \simeq 100 \textrm{ Hz}$ and
$\sqrt{ {\langle{ \vert{ \hat{x}_T (1 \textrm{ Hz})} \vert}^2
\rangle} } \simeq 10^{-11} \textrm{ m} / \sqrt{\mathrm{Hz}}$,
which is a very small signal. As a consequence, extremely small
vibrations of the environment may greatly perturb the measurement.
Therefore, to increase the signal-to-noise ratio of the
measurement, one has to reduce the coupling of the cantilever to
the environmental noise (acoustic and seismic) using vibration
isolation systems. This may be not enough in this specific case
because of the     smallness of the thermal fluctuations.
Therefore we have applied an original noise subtraction technique
described in ref.\cite{RSInoise} in order to recover $x_T$ from
the measurement of $x_c$.

The measurement of $x_c$ is done using a Nomarski interferometer
(for detailed reviews, see \cite{bib4, bib5, bib6}) which uses the
mirror glued on the Polycarbonate cantilever in one of the two
optical paths. The interferometer noise is about $5 \times
10^{-14} \textrm{ m} / \sqrt{\mathrm{Hz}}$, which is two orders of
magnitude smaller than the cantilever thermal fluctuations. The
cantilever is inside an oven under vacuum. A window allows the
laser beam to go inside (cf Fig.\,\ref{fig2}). The size of the
Polycarbonate cantilever are, $l \simeq 13.5 \textrm{ mm}$, $a
\simeq 1 \textrm{mm}$ and $b = 125\ \mu \mathrm{m}$, and the
mirror mass is $m_{\mathrm{mirror}} \lesssim 10^{-3} \textrm{ g}$
such that  $\nu_0 \simeq 100 \textrm{ Hz}$. As already mentioned,
the cantilever is sensitive to unavoidable mechanical vibrations
which are the main source of error and strongly reduce the signal
to noise ratio.  To improve the signal-to-noise ratio we have
applied     the reduction technique described in
ref.\cite{RSInoise}. This technique is based on a mechanical noise
detection system whose scheme is shown in fig.\ref{fig2} :
A second cantilever, the parameter of which are tuned to be only sensitive to
external vibration (and not to its own thermal fluctuations), is used to subtract
the mechanical noise component from the signal of the
polycarbonate cantilever.
More details can be found in ref.\cite{RSInoise}.

\begin{figure}
\centerline{\hspace{1cm} \bf (a) \hspace{8cm} (b) }
\begin{center}
\includegraphics[width=8cm]{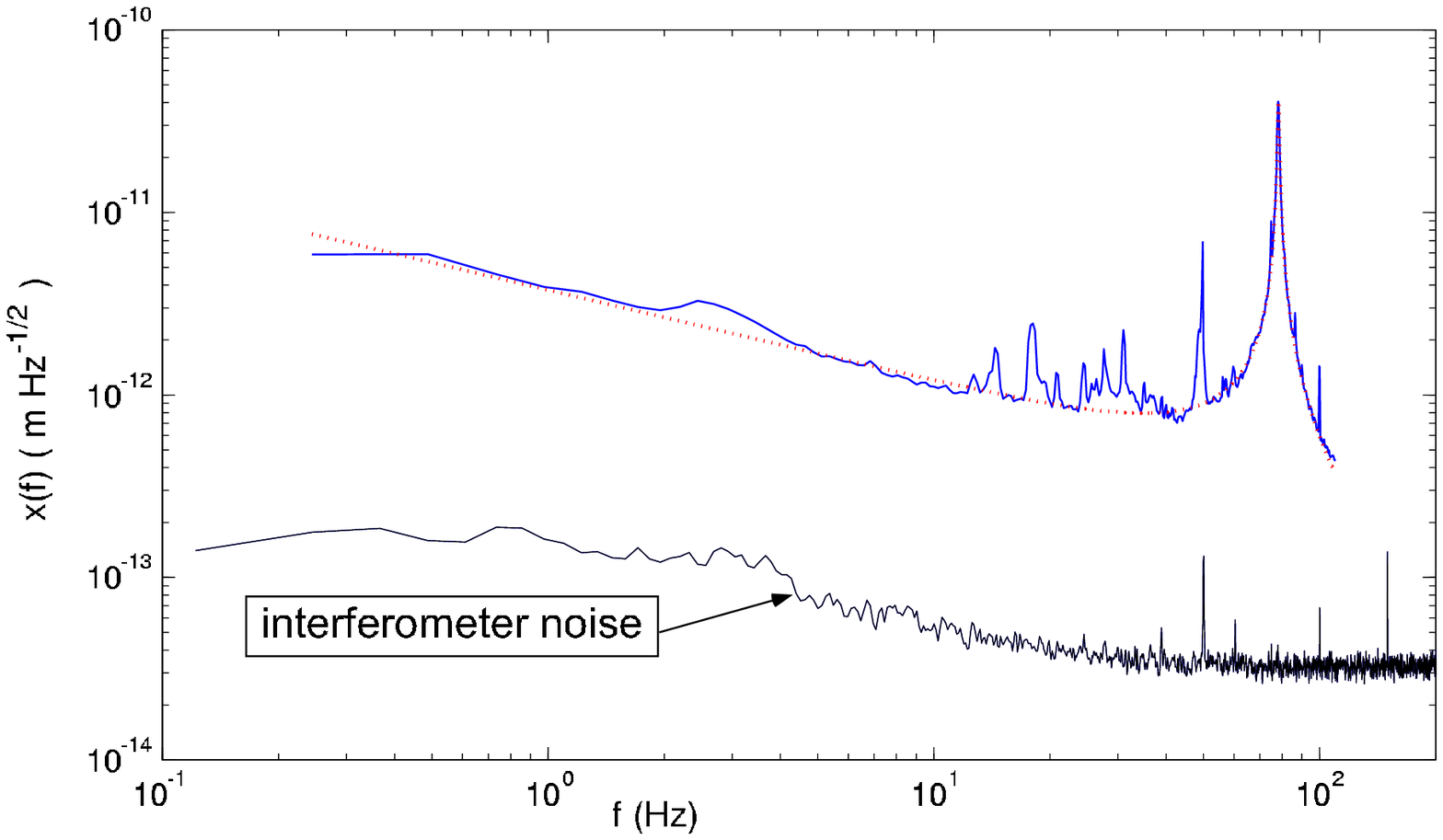}
\hspace{1mm}
\includegraphics[width=8cm]{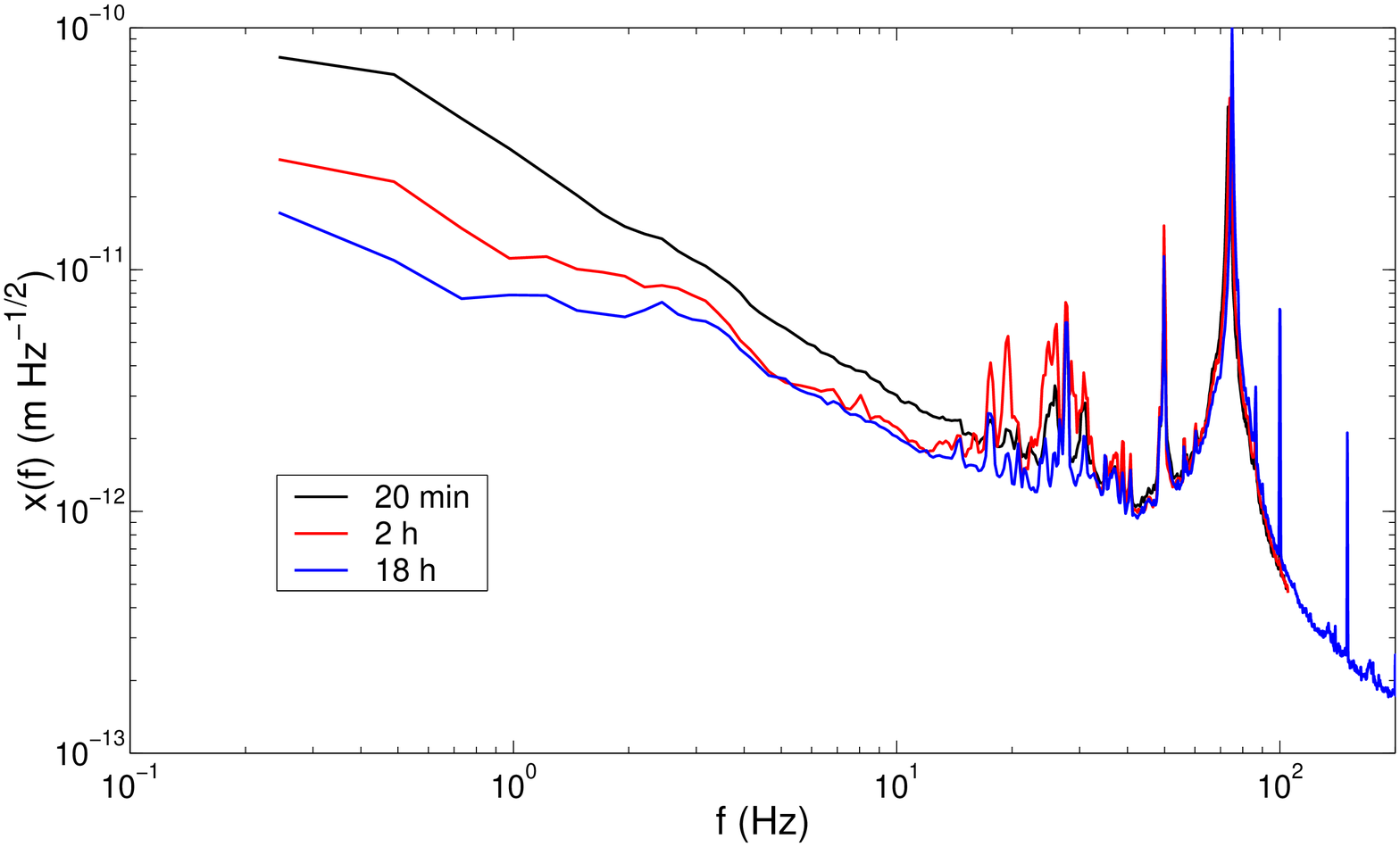}
\end{center}
    \caption{{\bf Spectra of the thermal  noise of the polycarbonate cantilever} (a) Equilibrium spectrum
    at 293K of the cantilever tip thermal fluctuations. The dashed line is the FDT prediction.
    the noise of the detection system (spectrum in the bottom of the figure)
     is shown for comparison. (b) Time evolution of the cantilever noise spectrum  recorded
     at three different times $tw=20 min,\ 2h, \ 18h$ after
     that the temperature $T_f=393K$ has been reached from below (see  text)}
     \label{FDTmech}
\end{figure}

\subsection{Experimental results} \label{sec4}

We first check whether the polycarbonate cantilever verifies the
FDT at room temperature. The results are shown in
fig.\ref{FDTmech}a), where the square root of the spectral density
is plotted as a function of f. The dashed line is the FDT
prediction obtained from a direct measurement of the cantilever
response function. The agreement is good. For comparison the
square root of the spectral density of the interferometer noise is
plotted too. We see that the SNR is quite good. The extra picks on
the spectrum of $x_c$ come from residual mechanical vibrations.
This figure shows that the experimental system is well suited to
study fluctuation relations in an aging material. To study the
polycarbonate cantilever noise after a quench we used a protocol
that is different from that described in section 1) for dielectric
measurement. As polycarbonate is almost liquid above $T_g$ it is
impossible to keep it in the above described measurement cell.
Thus we put the cantilever inside a box which has in the bottom a
groove where the cantilever fits perfectly inside. Then this box
is first heated at $T_i=460K$  and then rapidly cooled by putting
it into a cool water. Thus in a few seconds the cantilever is
quenched from $460K$  to about $280K$. Then the cantilever is installed inside the
 measurement cell that is heated to the working temperature $T_f$
 which is now reached from low temperature. The polycarbonate ages
  anyway:  it is well known   that in aging systems the time
 spent at low temperature does not affect the aging at high
 temperature. We report here a measurement of the time evolution of the noise spectrum
 performed at   $T_f=0.93T_g$.
A typical evolution of the   polycarbonate   cantilever
 fluctuation spectrum  is plotted in fig.\ref{FDTmech}. The spectra
 recorded at $t_w=20min, \ 2h, \ 18h$ are shown. We see that at very short time the
 spectrum present a very large power law behaviour at low
 frequencies. This component relaxes towards equilibrium in several
 hours. As the response function of the cantilever evolves of just
 a few percent during the same amount of time it is clear that the
 violation of FDT is very large also in the case of this mechanical
 measurement. The reason is the presence  of a strong
 intermittency as in the case of the dielectric measurements
 described in Sec.~2.

\section{Thermal noise in a colloidal glass}

We   review   in this section     results on electrical noise
measurements in Laponite during the transition from a fluid like
solution to a solid like colloidal glass. The main control
parameter of this transition is the concentration of
Laponite\cite{Laponite}, which is a synthetic clay consisting of
discoid charged particles. It disperses rapidly in water to give
gels even for very low mass fraction. Physical properties of this
preparation evolves for a long time, even after the sol-gel
transition, and have shown many similarities with standard glass
aging\cite{Kroon}. Recent experiments have even proved that the
structure function of Laponite at low concentration (less than $3
\%$ mass fraction) is close to that of a glass, suggesting the
{\it colloidal glass} appellation \cite{Bonn}.

In previous studies, we showed that the early stage of this
transition was associated with a small aging of its bulk
electrical conductivity, in contrast with a large variation in the
noise spectrum at low frequency. As a consequence, the FDT in this
material appeared to be strongly violated at low frequency in
young samples, and it is only fulfilled for high frequencies and
long times\cite{Bellon,BellonD,buisson,Buisson}. As in
polycarbonate, this effect was shown to arise from a strong
intermittency in the electrical noise of the samples,
characterized by a strong deviation to a standard gaussian
noise\cite{Buisson}. We summarize these results in the first part
of this section, before presenting preliminary results on the role
of concentration     in the noise behavior.

\subsection{Experimental setup}

The experimental setup is similar to that of previous
experiments\cite{Bellon,BellonD,Buisson}. The
Laponite\cite{Laponite} dispersion is used as a conductive
material between the two golden coated electrodes of a cell. It is
prepared in a clean $\mathrm{N_2}$ atmosphere to avoid
$\mathrm{CO_2}$ and $\mathrm{O_2}$ contamination, which perturbs
the aging of the preparation and the electrical measurements.
Laponite particles are dispersed at a concentration of $2.5 \%$ to
$3 \%$ mass fraction in pure water under vigorous stirring for
$300\,s$. To avoid the existence of any initial structure in the
sol, we pass the solution through a $1\,\mu m$ filter when filling
the cell. This instant defines the origin of the aging time $t_w$
(the filling of the cell takes roughly two minutes, which can be
considered the maximum inaccuracy of $t_w$). The sample is then
sealed so that no pollution or evaporation of the solvent can
occur. At these concentrations, light scattering experiments show
that Laponite\cite{Laponite} structure functions are still
evolving several hundreds hours after the preparation, and that
solid like structures are only visible after $100\,h$\cite{Kroon}.
We only study the beginning of this glass formation process.

\begin{figure} \begin{center}

 \null  \hspace{10mm}  (a) \hspace{46mm} (b)  \hspace{53mm}  (c)  \null

\includegraphics[scale=0.9]{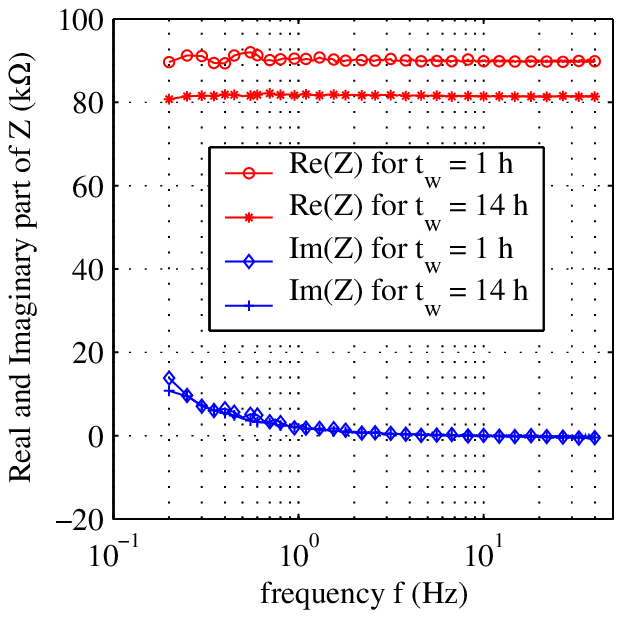}
\hfill
\includegraphics[scale=0.9]{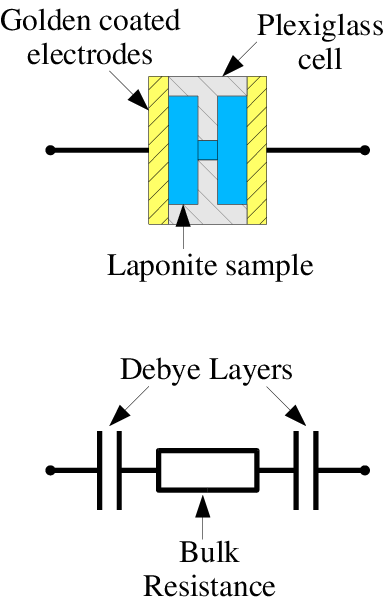}
\hfill
 \includegraphics[scale=0.9]{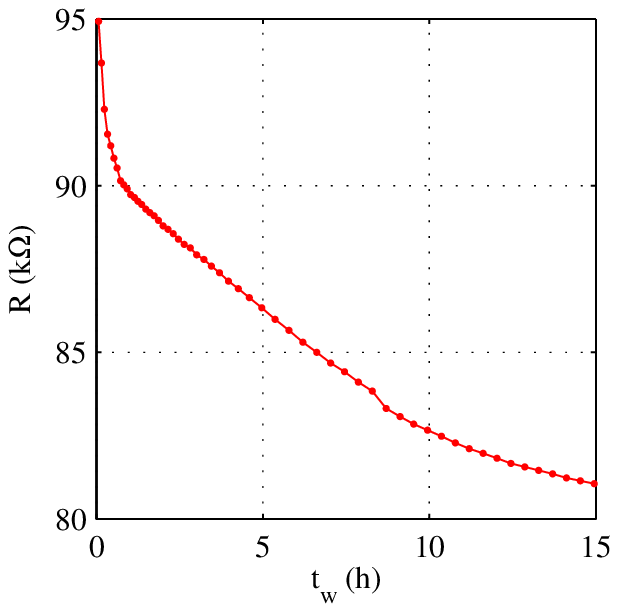}

 \end{center}
 \caption{{\bf Impedance of a $2.5\,wt\%$ Laponite cell. }(a) Frequency dependance
 of a sample impedance for 2 different aging times ($t_w=1\,h$ and $t_w=14\,h$).
 (b) Cell design and equivalent electrical model
 (c) Time evolution of the bulk resistance: this long
 time evolution is the signature of the aging of the
  colloidal suspension. In spite of the decreasing mobility of
   Laponite particles in solution during the formation of the gel,
    the electrical conductivity increases.}
 \label{fig:impedance}
\end{figure}

The two electrodes of the cell are connected to our measurement
system, which records either the impedance value or the voltage
noise across it. The electrical impedance of the sample is the sum
of two effects: the bulk is purely conductive, the ions of the
solution following the forcing field, whereas the interfaces
between the solution and the electrodes give mainly a capacitive
effect due to the presence of the Debye layers\cite{hunter}. This
behavior has been validated using a four-electrode potentiostatic
technique\cite{electrochem} to make sure that the capacitive
effect is only due to the surface. In order to probe mainly bulk
properties, the geometry of the cell is tuned to push the surface
contribution to low frequencies: the cell consists in two large
reservoirs where the fluid is in contact with the electrodes (area
of $25\,cm^2$), connected through a small rigid tube --- see
  Fig.~\ref{fig:impedance}(b).   The main contribution to the
electrical resistance of the cell is given by the Laponite sol
contained in this tube connecting the two tanks. Thus by changing
the length and the section of this tube the total bulk resistance
of the sample can be tuned around $R_{opt} = 100\,k\Omega$, which
optimizes the signal to noise ratio of voltage fluctuations
measurements with our amplifier. The cut-off frequency of the
equivalent R-C circuit (composed by the series of the Debye layers
plus the bulk resistance) is about $20 \,mHz$. In other words
above this frequency the imaginary part of the cell impedance is
about zero, as shown in Fig.~\ref{fig:impedance}(a). The time
evolution of the resistance of one of our sample is plotted in
Fig.~\ref{fig:impedance}(c): it is still decaying in a non trivial
way after $24 h$, showing that the sample has not reached any
equilibrium yet. This aging is consistent with that observed in
light scattering experiments\cite{Kroon}.

 \subsection{Electric noise measurements in Laponite}

\begin{figure}[tb]
\begin{center}
 \includegraphics[width=8cm]{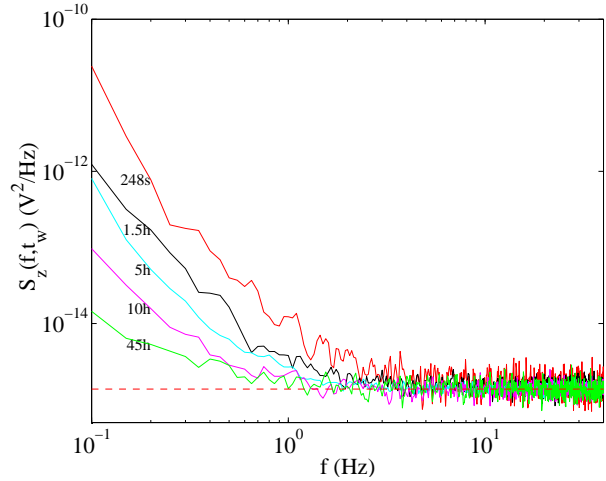}
 \end{center}
\caption{{\bf Voltage noise density for a $2.5\,wt\%$ Laponite
sample. }The power spectrum density of voltage fluctuations across
the impedance of a $2.5\,wt\%$ Laponite cell exhibit strong aging,
and match the Nyquist formula prediction (horizontal dashed line)
only for long times or large frequencies.} \label{fig:LaponitePSD}
\end{figure}

In order to study the voltage fluctuations across the Laponite
cell, we use a custom ultra low noise amplifier to raise the
signal level before acquisition. To bypass any offset problems
during this strong amplification process, passive high pass
filtering above $30\,mHz$ is applied. The power spectrum density
of the voltage noise of a $2.5\,wt\%$ Laponite preparation is
shown in Fig.~\ref{fig:LaponitePSD}. As the dissipative part of
the impedance $Re(Z)$ is weakly time and frequency dependent, one
would expect from the Nyquist formula\cite{Nyquist} that so does
the voltage noise density $S_{Z}$. But as shown in
Fig.~\ref{fig:LaponitePSD}, we have a large deviation from this
prediction for the lowest frequencies and earliest times of our
experiment: $S_{Z}$ changes by several orders of magnitude between
highest values and the high frequency tail. For long times and
high frequencies, the FDT holds and the voltage noise density is
that predicted from the Nyquist formula for a pure resistance at
room temperature ($300\,K$). In order to be sure that the observed
excess noise is not due to an artifact of the experimental
procedure, we filled the cell with an electrolyte solution with a
{\it p}H close to that of the Laponite preparation such that the
electrical impedance of the cell was the same (specifically:
$NaOH$ solution in water at a concentration of $10^{-3} \, mol
\cdot l^{-1}$). In this case, the noise spectrum was flat and in
perfect agreement with the Nyquist formula\cite{Bellon}.

\begin{figure}[!h]
 \begin{center}
 \includegraphics[width=7cm]{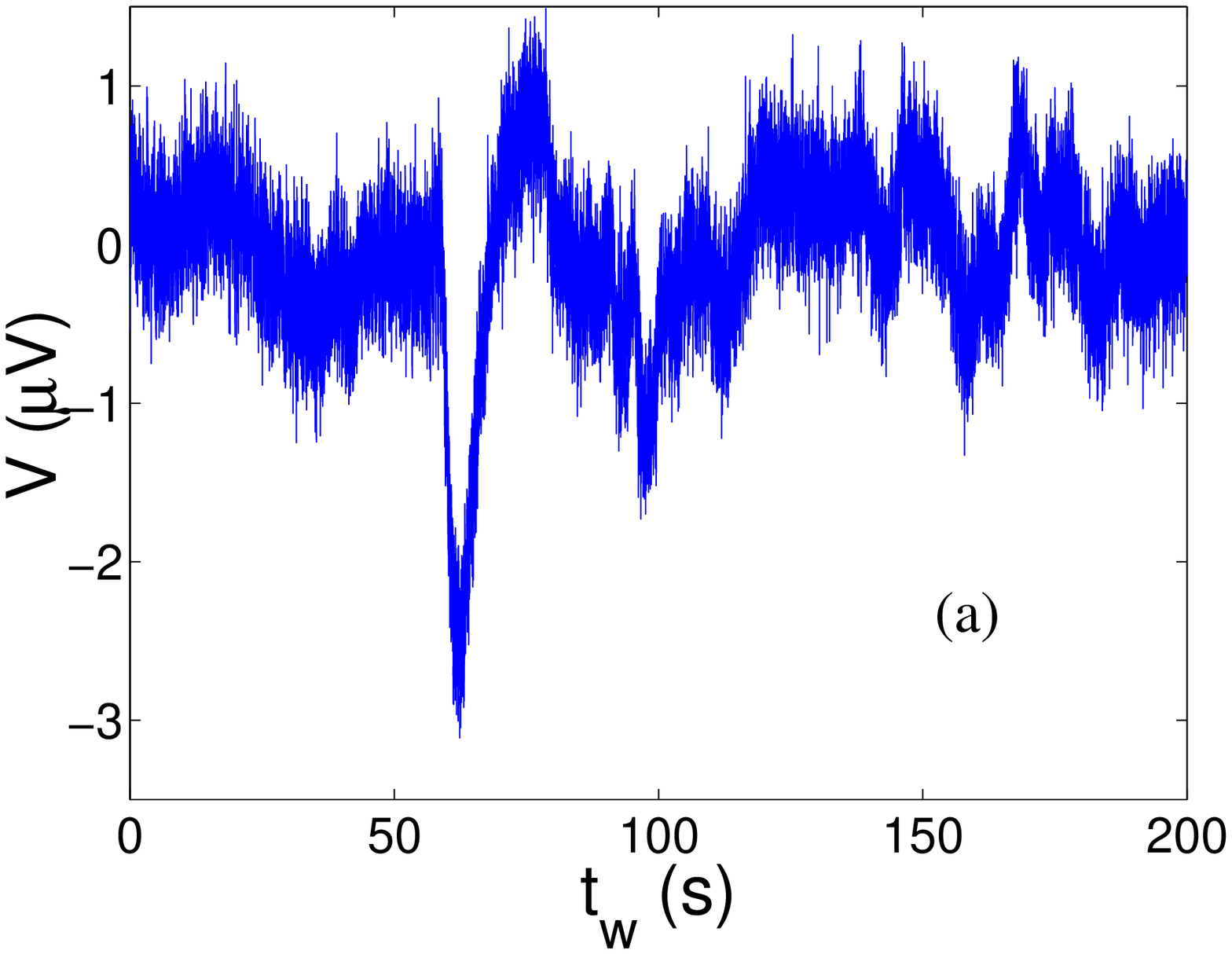}
 \hspace{15mm}
 \includegraphics[width=7cm]{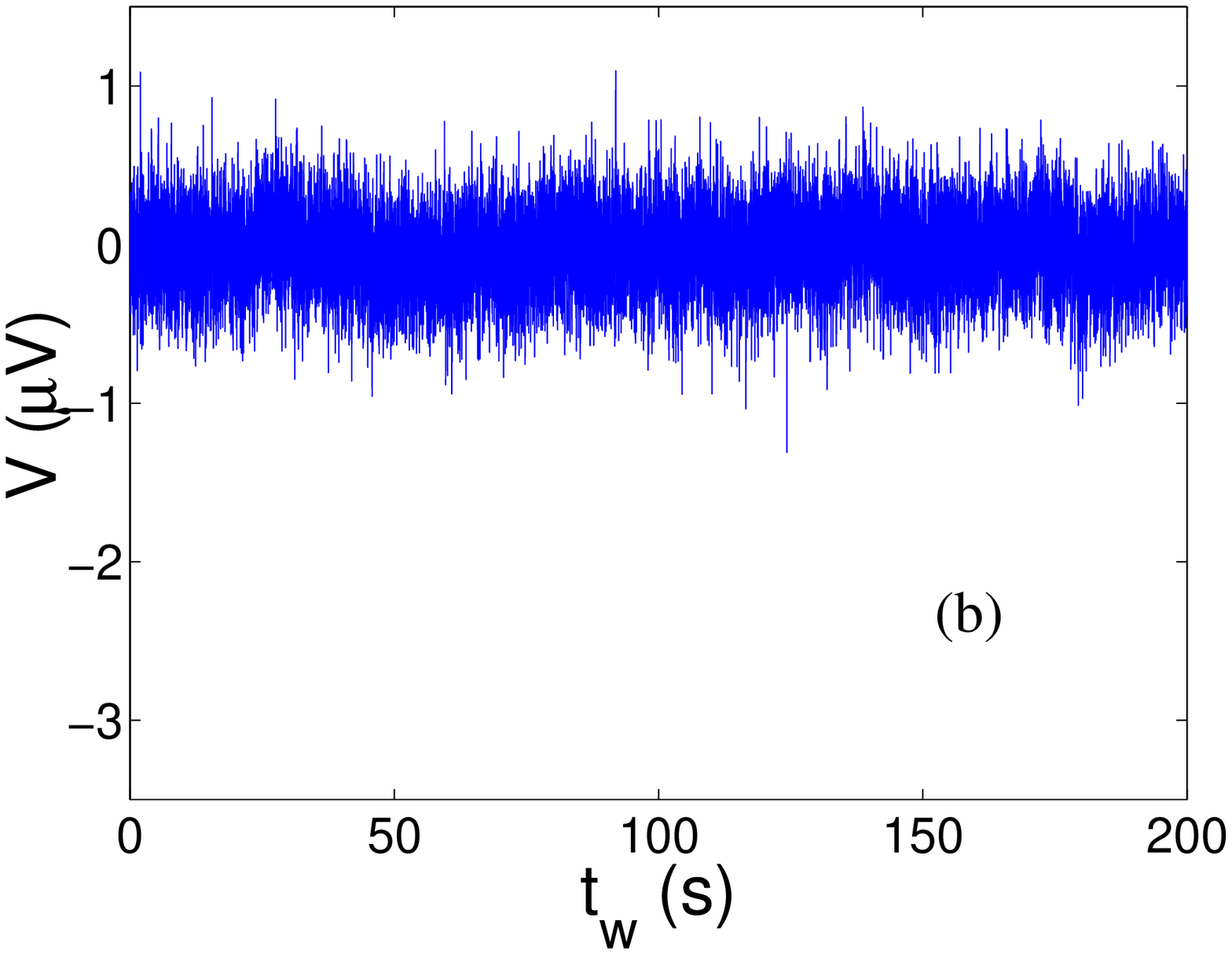}
 \end{center}
\caption{{\bf Voltage noise signal in a $2.5\,wt\%$ Laponite
sample. }(a) Noise signal, 2 hours after the Laponite preparation,
when FDT is violated. (b) Typical noise signal when FDT is not
violated ($t_w=50\,h$). } \label{fig:signal2.5}
\end{figure}

Aiming at a better understanding of the physics underlaying such a
behavior, we have directly analyzed the voltage noise across the
Laponite cell. This test can be safely done in our experimental
configuration as the amplifier noise is negligible with respect to
the voltage fluctuations across cell, even for the lowest levels
of the signal, that is when the FDT is satisfied. In
Fig.~\ref{fig:signal2.5}(a) we plot a typical signal measured $2h$
after the gel preparation, when the FDT is strongly violated. The
signal plotted in Fig.~\ref{fig:signal2.5}(b) has been measured
when the system has relaxed and FDT is satisfied in all the
frequency range. By comparing the two signals we immediately
realize that there are important differences. The signal in
Fig.~\ref{fig:signal2.5}(a) is interrupted by bursts of large
amplitude which are responsible for the increasing of the noise in
the low frequency spectra (see Fig.~\ref{fig:LaponitePSD}). The
relaxation time of the bursts has no particular meaning, because
it corresponds just to the characteristic time of the filter used
to eliminate the very low frequency trends. As time goes on, the
amplitude of the bursts reduces and the time between two
consecutive bursts becomes longer and longer. Finally they
disappear as can be seen in the signal of
Fig.~\ref{fig:signal2.5}(b) recorded for a $50\,h$ old
preparation, when the system satisfies FDT.

\begin{figure}[!h]
 \begin{center}
 \includegraphics[width=8cm]{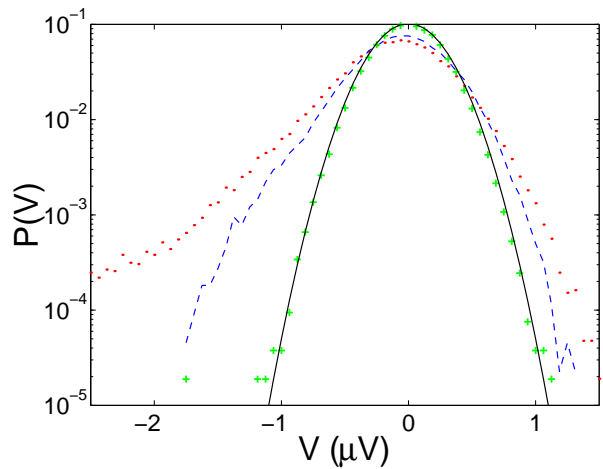}
 \end{center}
\caption{{\bf PDF of the voltage noise in a $2.5\,wt\%$ Laponite
sample. }Typical PDF of the noise signal at different times after
preparation, with from top to bottom: $
 \ (...) t_w=1\,h, \ (- -) t_w=2\,h, \ (+)
t_w=50\,h$. The continuous line is obtained from the FDT
prediction. } \label{fig:PDFlaponite2.5}
\end{figure}

As in polycarbonate, the intermittent properties of the noise can
be characterized by the PDF of the voltage fluctuations. To
compute these distributions, the time series are divided in
several time windows and the PDF are computed in each of these
window. Afterwards the result of several experiments are averaged.
The distributions computed at different times are plotted in
Fig.~\ref{fig:PDFlaponite2.5}. We see that at short $t_w$ the PDF
presents heavy tails which slowly disappear at longer $t_w$.
Finally a Gaussian shape is recovered after $t_w=16\,h $. This
kind of evolution of the PDF clearly indicate that the signal is
very intermittent for a young sample and it relaxes to the
Gaussian noise at long times.

\begin{figure}[!h]
\begin{center}
\includegraphics[width=7.02cm]{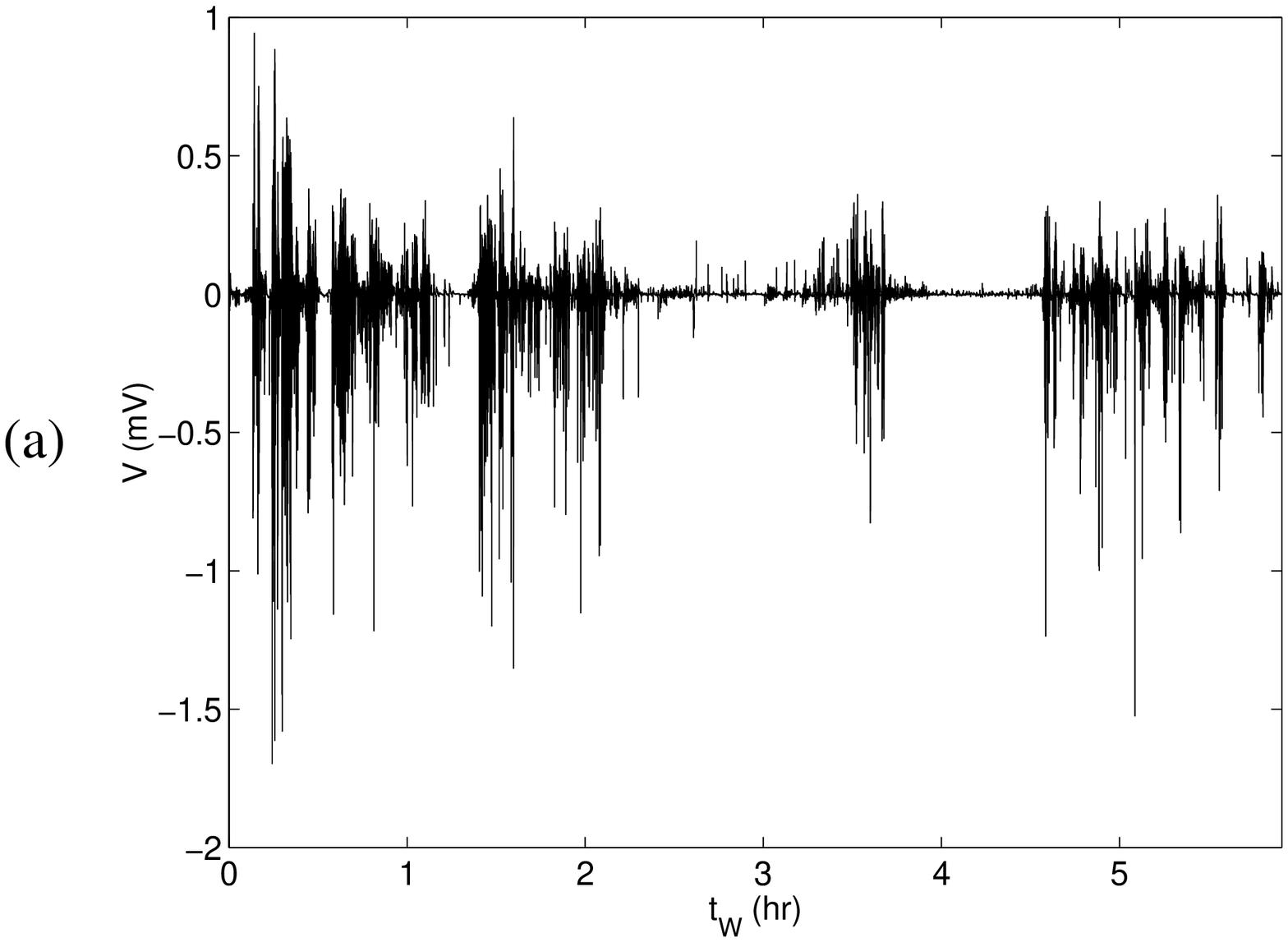} \hspace{1cm}
\includegraphics[width=7.02cm]{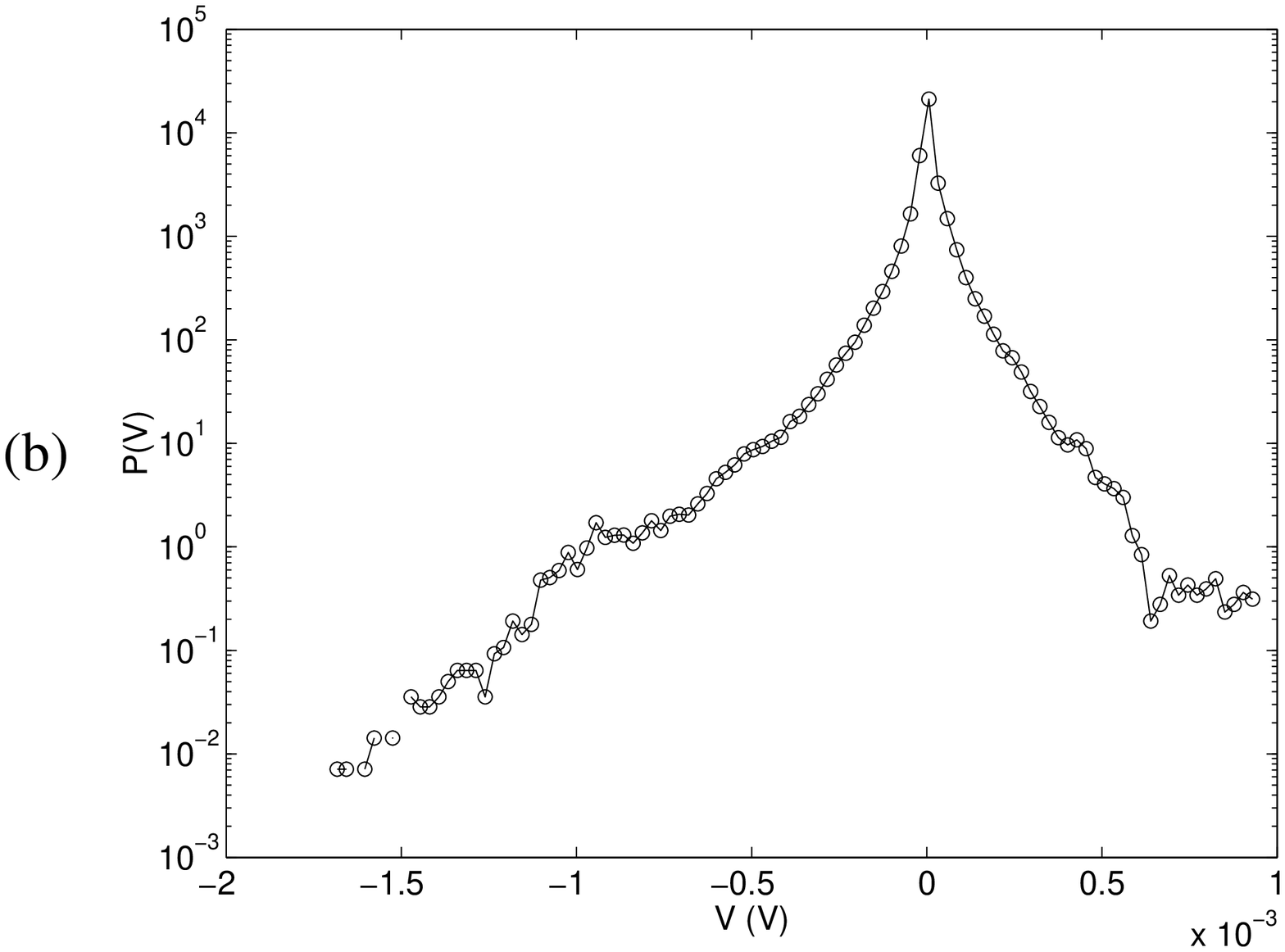}
\end{center}
\caption{{\bf Voltage noise in a $3\,wt\%$ Laponite sample. }(a)
During the first hours, the voltage noise is dominated by huge
intermittent fluctuations: bursts over $1\,mV $ are detected, when
thermal noise should present a typical $1\,\mu V$ rms. amplitude.
(b) The PDF of this intermittent signal departs clearly from a
gaussian distribution.} \label{fig:Laponite3Filtered}
\end{figure}

\subsection{Influence of concentration}

To check for the influence of concentration on these results, we
recently started new series of measurements with $3\,wt\%$
Laponite preparations. In Fig.~\ref{fig:Laponite3Filtered}(a) we
plot a typical signal measured during the first 6 hours of such a
sample. Again, this signal is interrupted by bursts of very large
amplitude. As time goes on, the amplitude of the bursts reduces
and the time between two consecutive bursts becomes longer and
longer. Finally they disappear after a few days, and we only
observe classic thermal noise. The main difference with less
concentrated samples is in the amplitude and density of this
intermittency: now bursts over $1\,mV $ are detected, when thermal
noise should present a typical $1\,\mu V$ rms. amplitude, and they
are much more frequent. This difference is also clear on the PDF
of the signal, plotted in Fig.~\ref{fig:Laponite3Filtered}(b). The
non gaussian shape is much more pronounced, and the presence of
heavy tails clearly indicate that the signal is very intermittent
at the beginning of the experiment. In fact, the dynamic is so
important that we don't even have enough precision to resolve the
classic thermal fluctuations predicted by the Nyquist formula in
this measurement. The influence of increasing the concentration of
Laponite preparation thus appears to be somehow similar to the
effect of increasing the cooling rate during the quench of the
polymer glass: the resulting dynamics is more intermittent in both
cases.

\subsection{ Mechanical noise on Laponite}

We have studied the mechanical noise of Laponite in very sensitive
thermal rheometer \cite{BellonD} which is based on a principle
very similar to the one described for the polycarbonate cantilever
in section 3)(see also \cite{Rheom}. We have found that in this
case no intermittency is present and the  violation of FDT, if it
exists, is certainly very small\cite{BellonD}. Recent measurements
done on the Brownian motion of a particle inside a Laponite
preparation   seems to confirm these observations \cite{Abou}.

\vspace{1cm}

\section{Discussion and conclusions}

In the previous sections we have presented several  measurements
of the electric and mechanical thermal noise in two very different
materials: a polymer and a colloidal glass. We first compare the
main results on the electric noise measurements which are
certainly the most complete. These results  are:
\begin{itemize}
\item[{\bf(1)}]  At the very beginning of aging   the noise amplitude for both materials
 is much  larger than what  predicted by Nyquist relations.
 In other words Nyquist relations, or more generally FDT, are violated because
 the materials are out of equilibrium: they are aging.
 In agreement with theoretical prediction the amplitude and the
persistence time of the FDT violation is a decreasing function of
frequency and time. The violation is observed even at $\omega t_w
\gg 1$ and it may last for more than $3h$ for $f>1Hz$.
\item[{\bf(2)}] The noise slowly relaxes to the usual value after a very long time.
\item[{\bf(3)}] For the polymer there is a large difference between fast and slow quenches.
In the first case the thermal signal is strongly intermittent, in
the second case this feature almost disappears. The features of
fast and slow quenches in polycarbonate are:
\begin{itemize}
  \item[{\bf(3.1)}]  {\bf After a fast quench} the $T_{eff}$ estimated using FDR is
  huge. This huge $T_{eff}$
  is produced by very large
intermittent bursts which are at the origin of the low frequency
power law decay of noise spectra. The statistic of these events is
strongly non Gaussian when FDT is violated and slowly relaxes to a
Gaussian one at very long $t_w$.  The time intervals $\tau$
between two intermittent events are power law distributed with an
exponent which depends on $T_f$.
  \item[{\bf(3.2)}] {\bf After a slow quench} the $T_{eff}$ estimated using FDR is about $20\%$ larger than $T_f$.
The intermittency disappears, the noise signal PDF are much closer
to a Gaussian and the time between two large fluctuations is not
power law distributed.
\end{itemize}
\item[{\bf(4)}] The colloidal suspension signal is strongly intermittent,
all the more as concentration is increased. The noise signal PDF
at small $t_w$ is strongly non-Gaussian. The asymmetry of the
noise may be linked to the spontaneous polarization of the cell.
 \end{itemize}

We want first to discuss the intermittence of the signal, which
has been observed in other aging systems. Our observations are
reminiscent of  the intermittence observed in the local
measurements of polymer dielectric properties
\cite{Israeloff_Nature} and in the slow relaxation dynamics of
another colloidal gel \cite{Mazoyer,Cipelletti}. Indeed several
theoretical models predict an intermittent dynamics for an aging
system. For example the trap model\cite{trap} which is based on a
phase space description of the aging dynamics. Its basic
ingredient is an activation process and aging is associated to the
fact that deeper and deeper valleys are reached as the system
evolves \cite{Sollich,Miguel1,Miguel2,Sibani,SibaniI}. The
dynamics in this model has to be intermittent because either
nothing moves or there is a jump between two traps\cite{Sollich}.
This contrasts, for example, with mean field dynamics which is
continuous in time\cite{Kurchan}. Furthermore two very recent
theoretical models predict skewed PDF both for local
\cite{Crisanti} and global variable \cite{SibaniI}. This is a very
important observation, because it is worth noticing that one
could expect to find intermittency in local variables but not in
global. Indeed in macroscopic measurements, fluctuations could be
smoothed by the volume average and therefore the PDF  would be
Gaussian. This is not the case both for our experiments and for the
numerical simulations of aging models\cite{SibaniI}. In order to
push the comparisons with these models of intermittency on a more
quantitative level one should analyze more carefully the PDF of
the time between events, which is very different in the various
models\cite{trap,Sibani,SibaniI}. Our statistics is not yet enough
accurate to give clear answers on this point, thus more
measurements are necessary to improve the comparisons between
theory and experiment. But the time statistics of the trap model
\cite{trap} seems to fit the data better than that of
\cite{Sibani}. The large   $T_{eff}$ produced by the intermittent
dynamics merits a special comment too. Indeed such a huge
$T_{eff}$ is not specific to this class of systems, it has also been observed
in domain growth models\cite{Peliti,Barrat}. The behaviour of
these models is however not consistent with that of our system, because in
the case of domain growth the huge temperature is given by a weak
response, not by an increase of the noise signal.

Going back to the analysis of our experimental data there is
another observation, which merits to be discussed. This concerns
the difference between fast and slow quenches in polycarbonate. In
order to discuss the problem related to this difference it is
important to recall that the zero of $t_w$ is defined as the
instant in which the temperature crosses $T_g$. The first question
that one may ask, already discussed in section 2.4, is whether the
behaviour of the system at the same $t_w$ after a slow and a fast
quenches is the same. This is certainly not the case because the
system takes about $20min$ during a slow quench to reach $T_f$ and
we have seen that after a fast quench the signal remains
intermittent for many hours, whereas after a slow quench
intermittency is never observed. Thus one concludes that it is not
just a matter of time delay between fast and slow quenches, the
dynamics is indeed very different in the two cases. This
result  can be understood considering that during the fast quench
the material is frozen in a state which is highly out of
equilibrium at the new temperature. This is not the case for
a slow quench. More precisely one may assume that when an aging
system is quenched very fast, it explores regions of its phase
space that are completely different than those explored in the
quasi-equilibrium states  of a slow quench. This assumption is
actually supported by two recent theoretical results
\cite{BertinKovacs,Sciortino}, which were obtained in order to
give a satisfactory explanation, in the framework of the more
recent models, of the old Kovacs effect on the volume
expansion\cite{Kovacs}. Our results based on the noise
measurements can be interpreted in the same way.

Finally, we want to discuss the analogy between the electrical
thermal noise  in the fast quench experiment of the polymer and
that of the gel during the sol-gel transition. In spite of the
physical mechanisms that are certainly very different, the
statistical properties of the signals are very similar. Thus one
may wonder, what is the relationship between the fast quench in the
polymer and the gel formation. As already mentioned,
during the fast quench the polymer is strongly out of equilibrium,
which is the same situation for the liquid-like state at the very
beginning of the gel transition. The speed of this
transition is controlled by the initial Laponite/water
concentration and therefore intermittency should be a function of
this parameter. Preliminary measurements seem to confirm this
guess: the higher the concentration, the stronger the
intermittency.

\begin{figure}[ht!]
\begin{center}
\includegraphics[width=8cm]{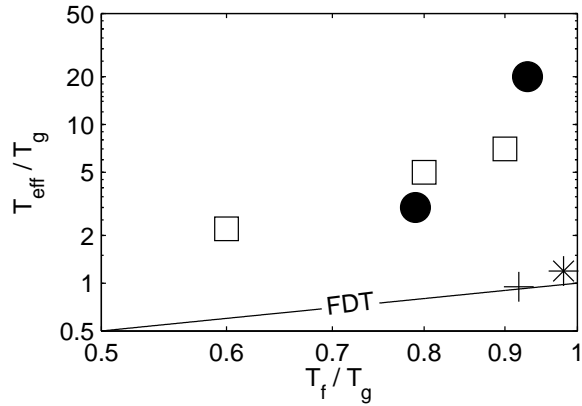}
\end{center}
\caption{{\bf $T_{eff}$ as a function of $T_f$}. $T_{eff}$
measured in several experiments on different types of glasses at
the beginning of the aging regime. (+) glycerol ($f=7Hz$)
\cite{Grigera}, ($\square$) spin glass ($q=q_{min}$) \cite{Herisson}, ($\bullet $)
polycarbonate ($f=7Hz$, fast quench), ($*$) polycarbonate ($f=7Hz$, slow quench)} \label{Teffvari}
\end{figure}

The main consequence of these observations in the electric
measurements is that the definition of $T_{eff}$ based on FDR
depends on the cooling rate (on the concentration for the colloid)
and probably on $T_f$. In fig.\ref{Teffvari} we have summarized
the $T_{eff}$  obtained  by electric measurements  performed on
glycerol\cite{Grigera} and on polycarbonate (Sec.2) and by
magnetic measurements performed on a spin glass \cite{Herisson}.
Specifically we plot $T_{eff}/T_g$ versus $T_f/T_g$. The straight
line is the FDT prediction for $T_{eff}$. Looking at this
figure we see that the situation is rather confused. However it
becomes more clear if one takes into account the cooling rate.  As
the $T_g$ is quite different in the various materials we define a
relative cooling rate $Q={\partial T \over
\partial t}{1\over T_g}$, which takes the following values: $0.5 \ min^{-1}$ for the
spin glass, $0.12 \ min^{-1}$ for the polycarbonate fast quenches
($T_f/T_g=0.93$ and $0.79$), $0.009 \ min^{-1}$ for the
polycarbonate slow quenches ($T_f/T_g=0.98$) and $0.012
\ min^{-1}$ for the glycerol experiment. Thus by considering
the relative cooling rate  it is clear that in the fast quenches
$T_{eff}$  is  very large and in the slow quenches it is small
independently of the material. However a dependence on $T_f$ seems
to be present too. Many  more measurements are certainly necessary
to confirm this dependence of $T_{eff}$ on $T_f$ and on the cooling
rate.

Let us now briefly discuss the results on the mechanical thermal
noise. To the best of our knowledge there are only three
measurements done on this kind of noise in aging systems, one  in
polycarbonate (Sec.3) and two in Laponite \cite{BellonD,Abou}. The
two measurements done in Laponite show that for short $t_w$ there
is no intermittency and the violation of FDT is very small. Thus
in the case of this colloidal glass different observables give
different $T_{eff}$. However this result contrasts with the one
described in Sec.3 where we have shown that the measurements of
the mechanical thermal noise agree with the electric ones because
after a fast quench both measurements confirm the presence of a
strong intermittency in the aging dynamics of polycarbonate. This
comparison between the mechanical and electric measurements in
polycarbonate is at the moment rather qualitative due to the
difficulty of the mechanical measurements. Much more precise data
are certainly necessary to give a clear answer. The difference
between the mechanical noise in polycarbonate and in Laponite is
still unexplained. It is certainly  related with the fact the
intermittency in the electrical measurements in Laponite  is
related to the  important role played by the  ions in the gel
formation.

We want to conclude by a few important and general questions which
remain opens. The first  concerns the quench rate. Indeed, is it
the speed in which $T_g$ is crossed that determines the dynamics
or the time in which $T_f$ is approached ? This question has been
already studied in the context of  response functions but it will
be important to analyze it in terms of noise.  The second
important open question  is why in realistic simulations of
Lenard-Jones glasses intermittency has not been observed
\cite{Kob1,Kob2}. Several hypothesis can be done:(i) The
simulations are done for a time which is too short to observe
intermittency which is a very slow phenomenon. (ii) In the
simulation the quench are performed at imposed volume, this is a
big difference with respect to the experiments  which are done at
imposed pressure. A third open question concerns the different
dynamics of the thermal noise measured on different observables.
Indeed even from a theoretical point view the effective
temperature of different observables  is the same in certain
models \cite{Berthier} and different in others \cite{Sollich}.
This is certainly a useful information that can give new insight
to the problem of the mechanisms of aging dynamics in different
materials.

This lecture clearly shows the importance of associating thermal
noise and response measurements. As we have already pointed out in
the introduction the standard techniques, based on response
measurements and on the application of thermal perturbations to
the sample, are certainly important to fix several constrains for
the phase space of the system. However they do no give
information on the dynamics of the sample, which can be obtained
by the study of FDR and of the fluctuation PDF.

{\noindent \bf Acknowledgments } We acknowledge useful discussion
with J.L. Barrat, J. P. Bouchaud, S. Franz and  J. Kurchan. We
thank P. Metz,   F. Vittoz  and D. Le Tourneau for technical
assistance. This work has been partially supported by the
DYGLAGEMEM contract of EEC, and by the contract ``Vieillissement
des mat\'eriaux amorphes'' of Région Rh\^one-Alpes.

\end{document}